\documentclass[manuscript,screen]{acmart}
\AtBeginDocument{%
  }

\setcopyright{acmlicensed}
\copyrightyear{2025}
\acmYear{2025}

\usepackage{tikz}
\usepackage{amsmath}

\usepackage{amsfonts}
\usepackage{algorithmic}
\usepackage{graphicx}
\usepackage{textcomp}
\usepackage{xcolor}
\usepackage{booktabs}
\usepackage{multirow}
\usepackage{array}
\usepackage{subcaption}
\usepackage{pifont}
\usepackage{url}
\usepackage{enumitem}

\usepackage{longtable}

\newcommand*\emptcirc[1][1ex]{\tikz\draw (0,0) circle (#1);} 
\newcommand*\halfcirc[1][1ex]{
	\begin{tikzpicture}
	\draw[fill] (0,0)-- (90:#1) arc (90:270:#1) -- cycle ;
	\draw (0,0) circle (#1);
	\end{tikzpicture}}
\newcommand*\fullcirc[1][1ex]{\tikz\fill (0,0) circle (#1);} 

\definecolor{darkgreen}{rgb}{0.0, 0.5, 0.0}
\definecolor{darkred}{rgb}{0.5, 0.0, 0.0}
\definecolor{darkblue}{rgb}{0.0, 0.24, 0.53}

\newcommand{\cmark}{\textcolor{darkgreen}{\ding{51}}}
\newcommand{\xmark}{\textcolor{darkred}{\ding{55}}}

\newcommand{\attackicon}{\includegraphics[width=0.3cm]{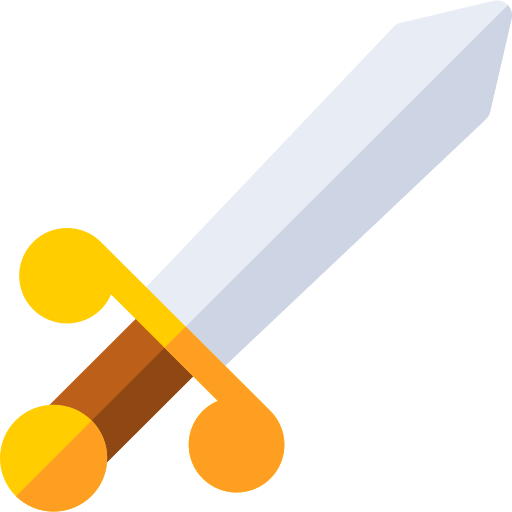}}
\newcommand{\defenseicon}{\includegraphics[width=0.3cm]{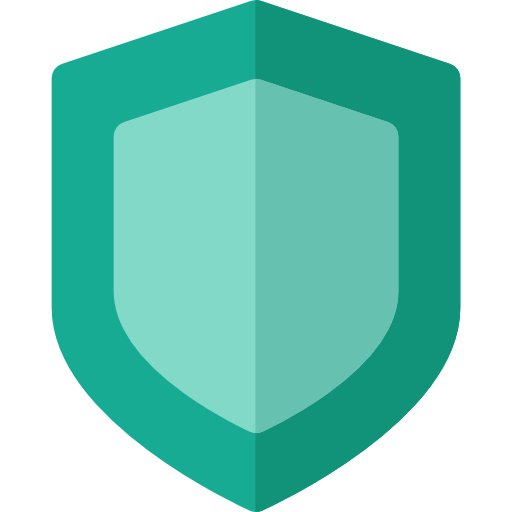}}

\newcommand{\apkicon}{\includegraphics[width=0.30cm]{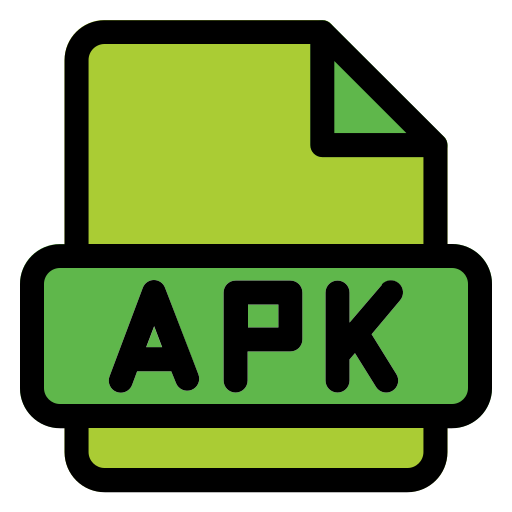}}
\newcommand{\exeicon}{\includegraphics[width=0.30cm]{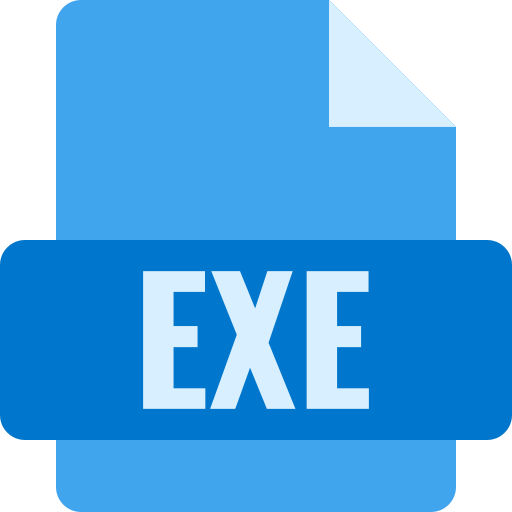}}
\newcommand{\pdficon}{\includegraphics[width=0.30cm]{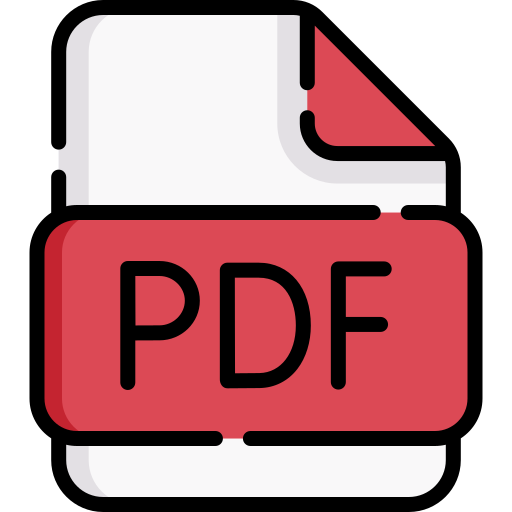}}

\newcolumntype{C}[1]{>{\centering\arraybackslash}m{#1}}

\newcommand{\paragraphbe}[1]{\noindent{\bf {#1}.}~}

\begin{document}

\title{On the Security Risks of ML-based Malware Detection Systems: A Survey}

\author{Ping He}
\email{gnip@zju.edu.cn}
\orcid{1234-5678-9012}
\affiliation{%
  \institution{Zhejiang University}
  \city{Hangzhou}
  \country{China}
}
\affiliation{%
  \institution{UCL}
  \city{London}
  \country{UK}
}
\author{Yuhao Mao}
\email{yuhao.mao@inf.ethz.ch}
\orcid{0009-0005-0520-9924}
\affiliation{%
  \institution{ETH Zürich}
  \city{Zürich}
  \country{Switzerland}
}
\author{Changjiang Li}
\email{meet.cjli@gmail.com}
\orcid{/0000-0002-1671-7183}
\affiliation{%
  \institution{Stony Brook University}
  \city{New York}
  \country{USA}
}
\author{Lorenzo Cavallaro}
\email{l.cavallaro@ucl.ac.uk}
\orcid{0000-0002-3878-2680}
\affiliation{%
  \institution{UCL}
  \city{London}
  \country{UK}
}
\author{Ting Wang}
\email{inbox.ting@gmail.com}
\orcid{0000-0003-4927-5833}
\affiliation{%
  \institution{Stony Brook University}
  \city{New York}
  \country{USA}
}
\author{Shouling Ji}
\email{sji@zju.edu.cn}
\orcid{0000-0003-4268-372X}
\affiliation{%
  \institution{Zhejiang University}
  \city{Hangzhou}
  \country{China}
}

\renewcommand{\shortauthors}{He et al.}

\begin{abstract}
Malware presents a persistent threat to user privacy and data integrity.
To combat this, machine learning-based (ML-based) malware detection (MD) systems have been developed.
However, these systems have increasingly been attacked in recent years, undermining their effectiveness in practice.
While the security risks associated with ML-based MD systems have garnered considerable attention, the majority of prior works is limited to adversarial malware examples, lacking a comprehensive analysis of practical security risks.
This paper addresses this gap by utilizing the CIA principles to define the scope of security risks.
We then deconstruct ML-based MD systems into distinct operational stages, thus developing a stage-based taxonomy.
Utilizing this taxonomy, we summarize the technical progress and discuss the gaps in the attack and defense proposals related to the ML-based MD systems within each stage.
Subsequently, we conduct two case studies, using both inter-stage and intra-stage analyses according to the stage-based taxonomy to provide new empirical insights.
Based on these analyses and insights, we suggest potential future directions from both inter-stage and intra-stage perspectives.
\end{abstract}




\maketitle

\section{Introduction}

Since the emergence of the Creeper Virus in 1971~\cite{Creeper}, malware has posed a persistent threat to computational devices. With over 1 billion identified malware samples~\cite{MalwareSample}, user privacy and data integrity are at significant risk.
In response, machine learning-based (ML-based) malware detection (MD) systems have been proposed and widely deployed in computational devices~\cite{KasperskyWP}.
These systems have gained popularity due to their effective use of powerful ML algorithms to efficiently detect and classify potential threats~\cite{AviraWP,AvastBG}.

In recent years, many attacks (e.g., StingRay~\cite{DBLP:conf/uss/SuciuMKDD18}, HRAT~\cite{DBLP:conf/ccs/ZhaoZZZZLYYL21}, JP~\cite{DBLP:conf/sp/YangCCPTPCW23}, BagAmmo~\cite{DBLP:conf/uss/0008CWYGYL23}, and AdvDroidZero~\cite{DBLP:conf/ccs/HeX0J23}) have targeted ML-based MD systems, compromising their ability to detect and protect against malware.
These attacks exploit vulnerabilities in program analysis and ML models.
To counter these emerging threatsss, various defenses~\cite{DBLP:journals/tdsc/DemontisMBMARCG19,DBLP:conf/www/LiZYLGC21,DBLP:conf/sp/BarberoPPC22,DBLP:conf/uss/LucasPLBRS23} have been proposed, enabling these systems to withstand security risks posed by dynamic environments and adversarial attackers.

Despite the growing interest in the security of ML-based MD systems, a comprehensive and systematic understanding of this field is still absent, hindering its further development.
Previous studies~\cite{DBLP:conf/uss/ArpQPWPWCR22,DBLP:conf/uss/PendleburyPJKC19,DBLP:conf/sp/RossowDGKPPBS12} have identified common pitfalls in developing such systems but have overlooked security threats from adversarial attackers, such as dataset poisoning risks.
In addition, recent surveys~\cite{DBLP:journals/csur/MaiorcaBG19,DBLP:journals/comsur/YanRWSZY23,DBLP:journals/csur/LiLYX23,DBLP:journals/compsec/LingWZQDCQWJLWW23} primarily focus on adversarial malware examples during system deployment, neglecting the complexities throughout the lifetime of an ML-based MD system.
In practice, it is crucial to recognize that security risks can emerge at any stage of the system's lifecycle, necessitating a comprehensive analysis framework to effectively assess and mitigate these risks.

To address the security risks in ML-based MD systems, we propose a comprehensive stage-based taxonomy that considers the entire lifecycle of these systems, dividing attack and defense proposals based on the four operational stages of the MD system. Through this taxonomy, we systematically examine the current proposals, identify research gaps, and enable system developers to understand and mitigate security risks at each stage. Our analysis offers a holistic perspective on attacks and defenses in ML-based MD systems, facilitating the development of more robust and secure systems.

We present our findings derived from the analysis of the existing attack and defense proposals in two directions: high-level findings (Section~\ref{sec:proposal_overview}) and low-level findings for each stage discussed as the open problems (Section~\ref{sec:sr1}-\ref{sec:sr4}).
High-level findings (in terms of CIA) include: 
1) although the security risks at each stage focus on different components of the ML-based MD system, they frequently target the integrity of the system, i.e., escaping detection;
2) confidentiality, i.e., stealing/protecting the copyrighted models and data, is systematically underrepresented in the existing proposals.
Low-level findings, to name a few, include:
1) strong knowledge assumptions and limited adaptability hinder the practicality of attacks in the malware domain, especially for backdoor attacks;
2) deobfuscation, as a popular defense, has limited practicality due to the necessary software or hardware requirements;
3) certified defense, which guarantees non-existence of certain attacks, is emerging but still exploratory in malware detection;
4) current adversarial query-based attacks merely insert dummy benign features into the malware and keep the malicious features intact, while the latter could be designed to ease the attack;
5) recent adaptive attacks have broken robust function call features which are the underlying assumption of many state-of-the-art (SOTA) defense proposals.

Further, we conduct two case studies focusing on inter-stage and intra-stage analyses to provide empirical insights (Section \ref{sec:casestudy}). Specifically, we find that poisoned models in the malware domain are not necessarily more vulnerable to adversarial attacks, whereas the opposite is true in the image domain~\cite{DBLP:conf/ccs/PangSZJVLLW20}.
This discrepancy is likely due to problem space constraints and the discrete nature of feature spaces in malware detection.
Additionally, concept drift detection methods in the malware domain exhibit a high true positive rate in identifying adversarial malware, likely due to the distribution shifts caused by adversarial perturbations.
We leave further exploration of these directions and findings to future work.
Finally, based on our taxonomy and analysis, we propose promising directions for future inter-stage and intra-stage research.

The main contributions of this work are summarized as follows.
We first establish the scope of the security risks by integrating the CIA principles with elements of ML-based MD systems.
Subsequently, we dissect the pipeline of the ML-based MD system into operational stages and propose a stage-based framework to holistically examine their security risks.
Based on the proposed framework, we systematize the technical progress and discuss the limitations of existing attack and defense proposals in ML-based MD systems at each stage.
Finally, we conduct two case studies from the perspective of both inter-stage and intra-stage analyses, resulting in novel insights.
Our analyses provide a comprehensive understanding of the security risks in ML-based MD systems and offer promising directions for future research.

\section{Background}


\subsection{Components of ML-based MD Systems}
\label{sec:mlmds}

MD systems are designed to detect software files exhibiting malicious behaviors, such as trojans, spyware, and ransomware.
Over the past two decades, the landscape of MD has seen considerable advances.
Traditionally, MD systems primarily rely on signature-based approaches, utilizing patterns and heuristics to identify malware.
These approaches necessitate a regularly updated malware signature database~\cite{DBLP:series/ais/27}.
However, such traditional systems are poorly generalized, especially in balancing the precision and recall of malware signatures.
To solve these challenges, SOTA MD systems employ ML techniques to enhance malware detection~\cite{DBLP:journals/jnca/GibertMP20,DBLP:journals/csur/LiuTLL23,DBLP:journals/jsa/SinghS21}.
Consequently, this research primarily focuses on ML-based MD systems.

\paragraphbe{Terminology} In this paper, we represent an ML-based MD system as $ \mathbb{F} = \{\mathbf{D}, \mathcal{F}, \mathcal{C} \} $, with the goal of classifying an unknown software sample $ u $ as either benign or malicious (i.e., detection result $ o $). The components are as follows:
\begin{itemize}

    \item \textit{File dataset} ($ \mathbf{D} $): The dataset used for training and evaluating ML-based MD systems. It is derived from a raw file database ($ \mathbf{\hat{D}} $) through a data preprocessing function ($ \mathcal{P} $), which selects the appropriate ratio and quantity of benign samples ($ x_{b} $) and malicious samples ($ x_{m} $).
    
    \item \textit{File feature extractor} ($ \mathcal{F} $): A mapping that converts file samples into feature vectors, representing the discernible characteristics of the file samples. Typically, three types of feature extraction methods are used: static analysis, dynamic analysis, and hybrid analysis.
    
    \item \textit{ML model} ($ \mathcal{C} $): A classifier that categorizes the features of a file, determining whether it is benign or malicious.
\end{itemize}

We remark that the terminology describing ML-based MD systems can vary across the literature.
The terms used in this paper are selected to facilitate a comprehensive analysis of the security risks associated with ML-based MD systems.

\begin{figure*}[t]   
	\centering  
	\includegraphics[width=1.0\linewidth]{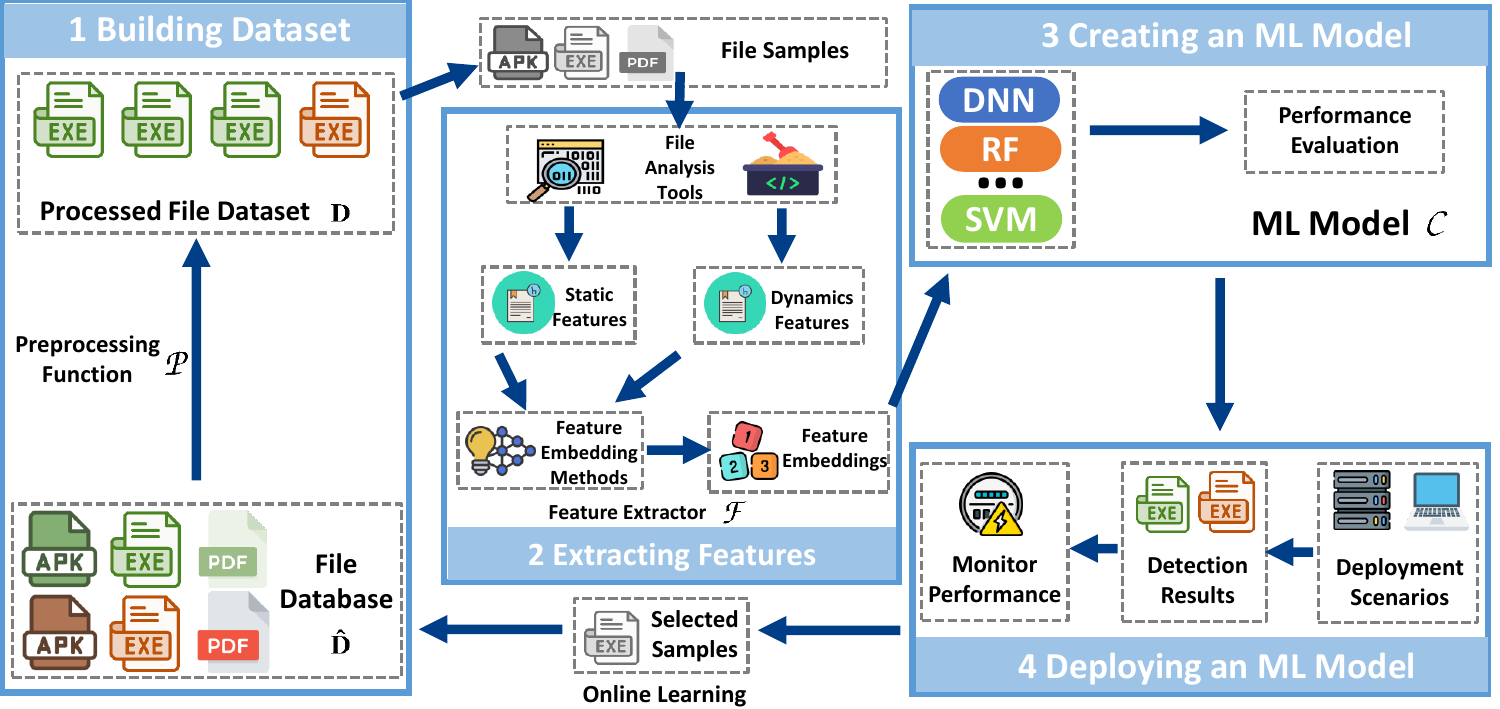}
	\caption{The pipeline of ML-based MD systems. It has four operational stages: (1) building a software file dataset, (2) extracting the software features, (3) creating an ML model, and (4) deploying an ML model.}
	\label{fig:mds}
\end{figure*}

\subsection{The Pipeline of ML-based MD Systems}
\label{sec:pipeline}

In this section, we divide the pipeline of ML-based MD systems into four operational stages from a practical lifecycle perspective.
These stages encompass both the development and deployment of ML-based MD systems, offering a comprehensive framework for analyzing security risks. Figure~\ref{fig:mds} illustrates the four operational stages of ML-based MD systems, denoted as $\mathbb{F} = \{\mathbf{D}, \mathcal{F}, \mathcal{C}\}$.

\paragraphbe{Stage 1: building a software file dataset $\mathbf{D}$ through $ \mathcal{P} $}
In contrast to benchmark datasets for the image and text domains, such as MNIST~\cite{DBLP:journals/pieee/LeCunBBH98}, ImageNet~\cite{DBLP:conf/cvpr/DengDSLL009}, IMDB Reviews~\cite{maas-EtAl:2011:ACL-HLT2011}, and Yelp Reviews~\cite{DBLP:conf/nips/ZhangZL15}, creating a software benchmark dataset presents unique challenges. Software data is highly dynamic due to frequent updates in programming languages, build tools, and associated libraries~\cite{DBLP:conf/uss/PendleburyPJKC19,DBLP:conf/aaai/RaffBSBCN18}.
Additionally, the computational cost of constructing a software file dataset is significantly higher than that for image or text datasets, as software files are often larger and require more computational and storage resources.
Furthermore, labeling software data is more costly, as it requires specialized domain knowledge\cite{aonzo2023humans}.
As a result, well-labeled software datasets are rare and valuable, thus usually regarded as proprietary~\cite{DBLP:journals/compsec/LingWZQDCQWJLWW23}.

The raw data used to construct the dataset primarily originates from online open-source software file repositories, such as VirusTotal~\cite{VirusTotal}, VirusShare~\cite{VirusShare}, and Google Play~\cite{GooglePlay}, or is privately collected by security products~\cite{DBLP:conf/uss/ShenVS22}.
We summarize the existing open-source datasets, as presented in Table~\ref{tab:MalDataset}.
Despite their utility, these datasets often contain extensive software files that exceed the processing capabilities of contemporary computing systems.
Moreover, the accessibility of software files from these sources tend to be limited and imbalanced, e.g., the BODMAS dataset~\cite{DBLP:conf/sp/YangCLA021} exclusively offers malicious samples for analysis.
Besides, recent studies~\cite{DBLP:conf/uss/PendleburyPJKC19,DBLP:conf/uss/ArpQPWPWCR22} have emphasized the importance of considering the practical constraints when creating software datasets for training and evaluation, including temporal consistency and space constraints, etc.
In this regard, an ML-based MD system $ \mathbb{F} $ utilizes a preprocessing function $ \mathcal{P} $ to strategically select the samples suitable for training and evaluation purposes.

\paragraphbe{Stage 2: extracting the software embedding features using $ \mathcal{F} $}
The software data in stage 1 is inherently file-based, characterized by complex structures and low-level semantics, presenting substantial challenges in training the ML model $ \mathcal{C} $. To enhance model training effectiveness, an ML-based MD system $ \mathbb{F} $ incorporates a dedicated software feature extractor $ \mathcal{F} $, which utilizes file analysis tools and/or feature embedding models to derive numerical embedding features from the software file data.

Static analysis methods are predominantly utilized in the feature extraction process due to their comparative efficiency over dynamic and hybrid analysis methods~\cite{DBLP:journals/csur/LiuTLL23,DBLP:conf/ccs/DambraHAKVCBB23}.
Static analysis examines the software file's structure and code without executing the file, providing a safer and quicker analysis. In contrast, dynamic analysis methods execute the software sample to monitor its runtime behavior, identifying suspicious function calls and cryptographic operations. Hybrid analysis combines static and dynamic approaches, offering a more comprehensive but more costly embedding~\cite{DBLP:journals/csur/LiuTLL23}.

The extraction process typically extracts features from three categories of software information: syntax information (e.g., permissions, strings, and file status), semantic information (e.g., function calls, data flow, and control flow), and raw bytes information (e.g., gray-scale images). These different types of information are extracted using various program analysis techniques. For instance, function calls can be identified through static analysis but may also be detected via dynamic analysis.

\paragraphbe{Stage 3: creating the ML model $ \mathcal{C} $}
The features extracted in stage 2 may have diverse types of software information. These features, due to their varied nature, are not tailored for a specific ML model $ \mathcal{C}$. For instance, syntax features, typically represented by binary values, are well-suited for an SVM model\cite{DBLP:conf/ndss/ArpSHGR14}.  Conversely, features based on function call graphs inherently possess graph structures and thus align better with graph-based models such as graph neural networks~\cite{DBLP:conf/infocom/LingWDQZZMWWJ22}. Therefore, an ML-based MD system $ \mathbb{F} $ may employ ensemble models to integrate heterogeneous features, with each child model processing a specific kind of feature\cite{DBLP:journals/tifs/KimKRSI19}. In cases where $ \mathbb{F} $ is designed with additional properties, more advanced ML model structures and training algorithms are employed. For example, \cite{DBLP:journals/tosem/WuCGFLWL21} applied the attention mechanism to encode interpretability, and \cite{DBLP:journals/tifs/FangHLLC0HZ23} applied federated learning to improve the effectiveness and preserve data privacy.

\paragraphbe{Stage 4: deploying the ML model $ \mathcal{C} $ at run-time}
At run-time, an ML-based MD system $ \mathbb{F} $ extracts features from the unknown software sample using $ \mathcal{F} $ and applies the trained ML model $ \mathcal{C} $ to classify the software sample. Once identified and confirmed as malware, the sample can be incorporated into the file database for subsequent analysis, such as exploring its relationship with other samples~\cite{VirusTotalGraph}.

\begin{table*}[t]\centering
    \setlength{\abovecaptionskip}{0pt}
	\caption{Open-source Malware Datasets. ``\emptcirc, \halfcirc, \fullcirc'' = only features are available, only malware files are available, all files are available.}
 
	\begin{tabular}[centering,width=0.5*\linewidth]{@{}C{2.5cm}C{1.5cm}C{1.5cm}C{0.8cm}C{1.5cm}C{1.5cm}@{}}
		\toprule
        Dataset & Type & Size & Time & Maintain & Availability \\
		\midrule
        VirusTotal~\cite{VirusTotal} & Hybrid & 3B+ & \cmark & \cmark & \halfcirc \\
        VirusShare~\cite{VirusShare} & Hybrid & 41M+ & \cmark & \cmark & \halfcirc \\
        AndroZoo~\cite{DBLP:conf/msr/AllixBKT16} & Andorid & 24M+ & \cmark & \cmark & \fullcirc \\
        Drebin~\cite{DBLP:conf/ndss/ArpSHGR14} & Andorid & 5,560 & \xmark & \xmark & \emptcirc \\
        MalRadar~\cite{DBLP:journals/pomacs/WangWHTMLL22} & Andorid & 4,534 & \cmark & \xmark & \halfcirc \\
        MARVIN~\cite{DBLP:conf/compsac/LindorferNP15} & Andorid & 150,000 & \xmark & \xmark & \fullcirc \\
        BODMAS~\cite{DBLP:conf/sp/YangCLA021} & WinPE & 134,435 & \cmark & \xmark & \halfcirc \\
        Ember~\cite{DBLP:journals/corr/abs-1804-04637} & WinPE & 1.1M & \cmark & \xmark & \emptcirc \\
        SOREL20M~\cite{DBLP:journals/corr/abs-2012-07634} & WinPE & 20M & \cmark & \xmark & \emptcirc \\
        Microsoft~\cite{DBLP:journals/corr/abs-1802-10135} & WinPE & 10,868 & \xmark & \xmark & \fullcirc \\
        \bottomrule
  	\end{tabular}
 
	\label{tab:MalDataset}
\end{table*}

\begin{table*}[t]\centering
    \setlength{\abovecaptionskip}{0pt}
	\caption{Representative ML-based MD systems according to the four stages.}
 
	\begin{tabular}[centering,width=0.5*\linewidth]{@{}C{2.5cm}C{2.0cm}C{1.5cm}C{1.5cm}C{1.5cm}C{1.5cm}C{1.8cm}@{}}
		\toprule
        \multirow{3}{*}{Proposals} & Stage 1 & \multicolumn{2}{c}{Stage 2} & Stage 3 & \multicolumn{2}{c}{Stage 4} \\
         \cmidrule(l){2-2} \cmidrule(l){3-4}  \cmidrule(l){5-5} \cmidrule(l){6-7}
        & Dataset Preprocessing & File Analysis & Feature Encoding & ML Model & Detection Task & Deploy Environment \\
		\midrule
        
        Drebin~\cite{DBLP:conf/ndss/ArpSHGR14} & None & Static  & Categorical & SVM & Malware Detection & Endpoint \\ \midrule
        Hidost~\cite{DBLP:conf/ndss/SrndicL13} & None & Static  & Tree & Decision Tree/SVM & Malware Detection & Endpoint \\ \midrule
        MaMadroid~\cite{DBLP:conf/ndss/MaricontiOACRS17} & None & Static & Numerical & Random Forest/kNN & Malware Detection & Server \\ \midrule
        MalwareExpert~\cite{DBLP:journals/tifs/ChenLHLH23} & Time-aware & Static  & Graph & GNN & Malware Detection & Server \\ \midrule
        MSDroid~\cite{DBLP:journals/tdsc/HeLWYRQ23} & None & Static & Graph & GNN & Malware Detection & Server \\ \midrule
        AndroAnalyzer~\cite{DBLP:journals/tifs/GongNLZZ24} & None & Static & Graph & GNN & Family Attribution & Server \\ \midrule
        MalConv~\cite{DBLP:conf/aaai/RaffBSBCN18} & None & Static & Bytes & CNN & Malware Detection & Server \\ \midrule
        Visual-AT~\cite{DBLP:journals/compsec/LiuLLZ20} & None & Static & Image & CNN & Malware Detection & Server \\ \midrule
        IMCEC~\cite{DBLP:journals/compsec/VasanAWSZ20} & None & Static & Image & CNN & Family Attribution & Server \\ \midrule
        LensDroid~\cite{DBLP:journals/tifs/MengZGWHCZX25} & None & Static & Hybrid & Hybrid Model & Malware Detection & Server \\ \midrule
        FEDRoid~\cite{DBLP:journals/tifs/FangHLLC0HZ23} & Random & Static & Categorical & ResNet & Malware Detection & Endpoint \\ \midrule
        Neurlux~\cite{DBLP:conf/acsac/JindalSALKV19} & Random & Dynamic & Behavior (Text) & CNN/LSTM & Malware Detection & Server \\ \midrule
        Rabadi \textit{et al}.~\cite{DBLP:conf/acsac/RabadiT20} & None & Dynamic & API Hash & Ensemble & Family Attribution & Server \\ \midrule
        APIChecker~\cite{DBLP:conf/eurosys/Gong00ZCQLL20} & None & Dynamic & Categorical (API) & Nine ML Models & Malware Detection & Server \\ \midrule
        XTrace+~\cite{DBLP:journals/tkde/ZhuCWXS24} & Random & Dynamic & Behavior (Text) & Transformer & Malware Detection & Server \\ \midrule
        Chaulagain \textit{et al}.~\cite{DBLP:conf/cns/ChaulagainPPRCL20} & None & Hybrid & Function Calls (Text) & LSTM & Malware Detection & Server \\
        \bottomrule
  	\end{tabular}
 
	\label{tab:MDWorks}
\end{table*}

In practice, the ML model $ \mathcal{C} $ may have diverse deployment scenarios. It can be deployed on user devices or centralized servers, representing on-device and off-device deployment cases, respectively. In the on-device deployment case~\cite{DBLP:journals/tifs/FengCXMLL21}, computational demands and user privacy concerns are critical considerations. Conversely, off-device deployment typically benefits from extensive computational resources and time budgets, allowing ML-based MD systems to utilize more complex ML models. Additionally, ML-based MD systems may employ a hybrid deployment strategy, operating as a distributed system consisting of both user-end devices and centralized servers. In this case, light computational tasks are executed locally, while intensive computations are transferred to remote servers.

Furthermore, the detection performance of an ML-based MD system is continuously monitored during deployment. As malware evolves and adversaries develop new variants capable of bypassing existing detection mechanisms, newly identified malware samples are selectively incorporated into the malware database to enhance detection capabilities. This process can be achieved through periodic retraining of the machine learning model or by employing online learning techniques, allowing the model to adapt to emerging threats in real-time.

Different ML-based MD systems adopt different methods for detecting different malware formats.
We summarize the existing ML-based MD systems in Table~\ref{tab:MDWorks} according to the methods used in four stages.

\section{Stage-Based Taxonomy}
\label{sec:taxonomy}

In this section, we first present the motivation behind the stage-based taxonomy.
Next, we establish the scope of security risks, utilizing the goals in the threat model.
Subsequently, we detail the strategies for each attack and defense proposal at each stage.
Lastly, an overview of the current field is provided according to our stage-based taxonomy.

\begin{table*}[]
    \centering
    \caption{CIA risks for each component of ML-based MD systems.} \label{tab:cia_risks}
    \begin{tabular}{ccccc}
        \toprule
          & $ \mathbf{D} $     & $ \mathcal{F} $             & $ \mathcal{C} $ & $ o $               \\ \midrule
        C & dataset leaking    & extraction function leaking & model leaking   & /                   \\
        I & dataset poisoning  & feature obfuscation         & model hijacking & high false negative \\
        A & dataset corruption & analysis blocking           & /               & high false positive \\ \bottomrule
    \end{tabular}
\end{table*}

\subsection{Motivation}
\label{sec:motivation}

In practice, the adversary often has multifaceted objectives encompassing all aspects of the CIA principles.
Additionally, a single system may be targeted by multiple attackers concurrently.
This complexity implies that the attacks against ML-based MD systems can occur at any stage in their operational pipeline and may involve various attacks simultaneously.
Conversely, defenders aim to secure the system comprehensively against all attacks at all stages.
Nevertheless, current surveys~\cite{DBLP:journals/compsec/LingWZQDCQWJLWW23,DBLP:journals/comsur/YanRWSZY23} in this field predominantly utilize taxonomy based on threat models, which fall short in capturing these practical complexities.
More discussion of the related surveys can be found in Section~\ref{sec:relatedwork}.

Recognizing the necessity for a more comprehensive understanding, we propose a holistic stage-based taxonomy utilizing the pipeline of ML-based MD systems.
This stage-based taxonomy is from the perspective of the ML-based MD system's developer, who aims to understand the security risks.
To reduce ambiguity, we identify the stage by the earliest time an attack influences the system, as developers prefer to eliminate risks at their entry point.
Although some attacks span multiple stages to take effect, our taxonomy decouples security risks for the developer, because each attack has an initial contaminated stage. For example, while dataset poisoning spans all stages to finally affect the detection results, we identify it as a Stage 1 attack, as it initially contaminates the dataset.
Hence, this stage-based taxonomy allows for effective assessment of potential security risks at each stage and investigation of interconnections between different stages in terms of attacks and defenses.

\subsection{Security Risk Scopes}
\label{sec:scope}

We establish the scope of the security risks in ML-based MD systems by referencing the goals within the threat model.
These goals pertain directly to the triad of information security principles: \textbf{confidentiality}, \textbf{integrity}, and \textbf{availability}, collectively known as the CIA principles.
As discussed in Section~\ref{sec:mlmds}, an ML-based MD system comprises three main elements: the file dataset $ \mathbf{D} $, the file feature extractor $ \mathcal{F} $, and the ML model $ \mathcal{C} $, along with the detection results $ o $.
Thus, the adversary's goals are undermining these principles within the three components and the detection results, while the defender's goals are to uphold them.
For instance, during deployment (Stage 4), the confidentiality of the ML model can be compromised by model extraction attacks, and the integrity of the detection results can be compromised by adversarial attacks. Correspondingly, defenders can preserve confidentiality by employing model obfuscation techniques and maintain integrity by improving the robustness of ML-based MD systems. We define the concrete goals of CIA principles at each stage in Section~\ref{sec:strategy} and further categorize all current attack and defense proposals according to these goals in Section~\ref{sec:proposal_overview}.

\begin{longtable}{@{}C{0.8cm}C{2.9cm}C{0.8cm}C{0.9cm}C{0.6cm}C{0.6cm}C{0.6cm}C{0.6cm}C{0.6cm}C{0.6cm}C{1.0cm}C{1.2cm}@{}}
\caption{Taxonomy of representative attack and defense proposals. ``Know. Assu.'' = Knowledge Assumption. ``PK, LK, ZK'' = Perfect Knowledge, Limited Knowledge, Zero Knowledge. ``Con., Int., Ava.'' = Confidentiality, Integrity, Availability. ``\attackicon, \defenseicon'' = Attack, Defense. ``\apkicon, \exeicon, \pdficon'' = Android, Windows PE, PDF.}
\label{tab:representativeworks} \\

\toprule
\multirow{3}{*}{\begin{tabular}[c]{@{}c@{}} Stage \end{tabular}} &
\multirow{3}{*}{\begin{tabular}[c]{@{}c@{}} Proposal \end{tabular}} &
\multirow{3}{*}{\begin{tabular}[c]{@{}c@{}}Year\end{tabular}} &
\multicolumn{9}{c}{Proposal Properties} \\ \cmidrule(l){4-12}
& & & \multirow{2}{*}{\begin{tabular}[c]{@{}c@{}}Know. \\ Assu. \end{tabular}} & \multicolumn{4}{c}{CIA Principles} & \multirow{2}{*}{\begin{tabular}[c]{@{}c@{}}Open \\ Source \end{tabular}} &  \multirow{2}{*}{\begin{tabular}[c]{@{}c@{}}	Real \\ World  \end{tabular}} &	\multirow{2}{*}{\begin{tabular}[c]{@{}c@{}}	Proposal \\ Type \end{tabular}}	& \multirow{2}{*}{\begin{tabular}[c]{@{}c@{}} Malware \\ Format \end{tabular}} \\ \cmidrule(l){5-8}
& & & & $ \mathbf{D} $  &  $ \mathcal{F} $  & $ \mathcal{C} $ & $ o $ &  &   &   \\ 

\midrule
\endfirsthead

\caption{(continued)} \\ 
\toprule
\multirow{3}{*}{\begin{tabular}[c]{@{}c@{}} Stage \end{tabular}} &
\multirow{3}{*}{\begin{tabular}[c]{@{}c@{}} Proposal \end{tabular}} &
\multirow{3}{*}{\begin{tabular}[c]{@{}c@{}}Year\end{tabular}} &
\multicolumn{9}{c}{Proposal Properties} \\ \cmidrule(l){4-12}
& & & \multirow{2}{*}{\begin{tabular}[c]{@{}c@{}}Know. \\ Assu. \end{tabular}} & \multicolumn{4}{c}{CIA Principles} & \multirow{2}{*}{\begin{tabular}[c]{@{}c@{}}Open \\ Source \end{tabular}} &  \multirow{2}{*}{\begin{tabular}[c]{@{}c@{}}	Real \\ World  \end{tabular}} &	\multirow{2}{*}{\begin{tabular}[c]{@{}c@{}}	Proposal \\ Type \end{tabular}}	& \multirow{2}{*}{\begin{tabular}[c]{@{}c@{}} Malware \\ Format \end{tabular}} \\ \cmidrule(l){5-8}
& & & & $ \mathbf{D} $  &  $ \mathcal{F} $  & $ \mathcal{C} $ & $ o $ &   &   &    \\ 
\midrule
\endhead

\midrule
\multicolumn{12}{r}{\small\itshape Continued on next page\ldots} \\
\endfoot

\bottomrule
\endlastfoot

\multirow{6}{*}{\begin{tabular}[c]{@{}c@{}} \MakeUppercase{\romannumeral 1} \end{tabular}} & Biggio \textit{et al}.~\cite{DBLP:conf/ccs/BiggioRAWCGR14} & 2014	& PK & Int. & \textbf{---} & \textbf{---} & Ava. & \xmark	& \xmark	& \attackicon	& \apkicon \exeicon \pdficon	\\
            & StingRay~\cite{DBLP:conf/uss/SuciuMKDD18} & 2018	& LK & Int. & \textbf{---} & \textbf{---} & Int.  & \cmark	& \xmark 	& \attackicon	& \apkicon	\\
		& Severi \textit{et al}.~\cite{DBLP:conf/uss/SeveriMCO21} & 2021	& PK & Int. & \textbf{---} & \textbf{---} & Int. & \cmark	& \xmark 	& \attackicon	& \apkicon \exeicon \pdficon	\\
		& Li \textit{et al}.~\cite{DBLP:journals/tdsc/LiCWWACX22} & 2022  		& PK & Int. & \textbf{---} & \textbf{---} & Int. & \xmark	& \xmark & \attackicon	& \apkicon	\\
		& JP~\cite{DBLP:conf/sp/YangCCPTPCW23} & 2023  	& LK & Int. & \textbf{---} & \textbf{---} & Int. & \cmark	& \xmark 	& \attackicon	& \apkicon	\\ 
        & SBV~\cite{DBLP:conf/uss/TianQG0KZ23} & 2023  	& LK & Int. & \textbf{---} & \textbf{---} & Int. & \cmark	& \xmark 	& \attackicon	& \apkicon \exeicon \pdficon	\\ 
        
		\midrule
  
		& Moser \textit{et al}.~\cite{DBLP:conf/acsac/MoserKK07} & 2007  	& ZK & \textbf{---} & Ava. & \textbf{---} & Int. & \xmark	& \cmark  & \attackicon	& \apkicon \exeicon \\
        & DroidChameleon~\cite{DBLP:journals/tifs/RastogiCJ14} & 2014  	& ZK & \textbf{---} & Int. & \textbf{---} & Int. & \xmark	& \cmark  & \attackicon	& \apkicon	\\
        & DexHunter~\cite{DBLP:conf/esorics/ZhangLY15} & 2015  	& PK & \textbf{---} & Int. & \textbf{---} &  \textbf{---} & \xmark	& \xmark  & \defenseicon	& \apkicon	\\
        & Mystique~\cite{DBLP:conf/ccs/MengXCN0ZC16} & 2016  	& ZK & \textbf{---} & Int. & \textbf{---} & Int. & \xmark	& \cmark & \attackicon	& \apkicon	\\
        & Mystique-S~\cite{DBLP:journals/tifs/XueM0TC0Z17} & 2017  	& ZK & \textbf{---} & Int. & \textbf{---} & Int.  & \xmark	& \cmark & \attackicon	& \apkicon	\\
        & MRV~\cite{DBLP:conf/acsac/YangK0G17} & 2017 	& ZK & \textbf{---} & Int. & \textbf{---} & Int. & \xmark	& \cmark	& \attackicon	& \apkicon	\\
        & SecureDroid~\cite{DBLP:conf/acsac/ChenHY17} & 2017 	& PK & \textbf{---} & Int. & \textbf{---}	& Int. & \xmark	& \xmark 	& \defenseicon	& \apkicon	\\
        \MakeUppercase{\romannumeral 2} & Malton~\cite{DBLP:conf/uss/XueZCLG17} & 2017 	& PK & \textbf{---} & Ava. & \textbf{---}	& Int. & \xmark	& \xmark	& \defenseicon	& \apkicon	\\
        & Hammad \textit{et al}.~\cite{DBLP:conf/icse/HammadGM18} & 2018  	& ZK & \textbf{---} & Int. & \textbf{---}	& Int. & \xmark	& \cmark & \attackicon	& \apkicon	\\
		& RevealDroid~\cite{DBLP:journals/tosem/GarciaHM18} & 2018  	& PK & \textbf{---} & Ava. & \textbf{---}	& Int. & \xmark	& \xmark	& \defenseicon	& \apkicon	\\
		& FARM~\cite{DBLP:journals/tifs/HanSX20} & 2020  	& PK & \textbf{---} & Int. & \textbf{---}	& Int. & \xmark	& \xmark	& \defenseicon	& \apkicon	\\
  	& API-Xray~\cite{DBLP:conf/uss/Cheng0LZFPM21} & 2021  	& PK & \textbf{---} & Int. & \textbf{---}	& Int. & \xmark	& \xmark	& \defenseicon	& \exeicon	\\
        & MalGraph~\cite{DBLP:conf/infocom/LingWDQZZMWWJ22} & 2022 & PK & \textbf{---} & Int. & Int.	& Int. & \cmark	& \xmark 	& \defenseicon	& \exeicon	\\
        & RLOBF~\cite{DBLP:journals/tdsc/WangWXZL22} & 2022 	& ZK & \textbf{---} & Int. & \textbf{---} & Int. & \cmark	& \cmark	& \attackicon	& \exeicon	\\
		& CorDroid~\cite{DBLP:journals/tifs/GaoCYHLYL23} & 2023	& PK & \textbf{---} & Ava. & \textbf{---}	& Int. & \xmark	& \xmark 	& \defenseicon	& \apkicon	\\
  
        \midrule
        
		\multirow{10}{*}{\begin{tabular}[c]{@{}c@{}} \MakeUppercase{\romannumeral 3} \end{tabular}} & Smutz \textit{et al}.~\cite{DBLP:conf/ndss/SmutzS16} & 2016  & PK & \textbf{---} & \textbf{---} & Int.	& Int. & \xmark & \xmark 	& \defenseicon	& \pdficon	\\
		& Wang \textit{et al}.~\cite{DBLP:conf/kdd/WangGZOXLG17} & 2017 	& PK & \textbf{---} & Int. & Int.	&  Int.	& \xmark & \xmark	& \defenseicon	& \exeicon	\\
		& Sec-SVM~\cite{DBLP:journals/tdsc/DemontisMBMARCG19} & 2017  	& PK & \textbf{---} & \textbf{---} & Int. & Int. & \xmark	& \xmark 	& \defenseicon	& \apkicon	\\
		& KuafuDet~\cite{DBLP:journals/compsec/ChenXFHXZL18} & 2018  	& PK & \textbf{---} & \textbf{---} & Int. & Int. & \xmark	& \xmark 	& \defenseicon	& \apkicon	\\
		& Fleshman \textit{et al}.~\cite{DBLP:journals/corr/abs-1806-06108} & 2018  & PK & \textbf{---} & \textbf{---} & Int. & Int. & \xmark	& \xmark	& \defenseicon	& \exeicon	\\
        & Chen \textit{et al}.~\cite{DBLP:conf/uss/Chen0SJ20} & 2020  & PK & \textbf{---} & \textbf{---} & Int. & Int. & \cmark	& \xmark & \defenseicon	& \pdficon	\\
		& Li \textit{et al}.~\cite{DBLP:conf/www/LiZYLGC21} & 2021  & PK & \textbf{---} & \textbf{---} & Int.	& Int. &  \cmark & \xmark	& \defenseicon	& \apkicon	\\
		& Lucas \textit{et al}.~\cite{DBLP:conf/uss/LucasPLBRS23} & 2023 & PK & \textbf{---} & \textbf{---} & Int. & Int. & \xmark & \xmark 	& \defenseicon	& \exeicon	\\
		& PAD~\cite{DBLP:journals/corr/abs-2302-11328} & 2023  	& PK & \textbf{---} & \textbf{---} & Int.	& Int. & \xmark & \xmark 	& \defenseicon	& \apkicon	\\
        & Gibert \textit{et al}.~\cite{DBLP:conf/ccs/GibertZL23} & 2023  	& PK & \textbf{---} & \textbf{---} & Int. & Int. & \cmark	& \xmark 	& \defenseicon	& \exeicon	\\
        & RS-Del~\cite{huang2023rs} & 2023 & PK & \textbf{---} & \textbf{---} & Int. & Int. & \cmark	& \xmark & \defenseicon	& \exeicon	\\
        & MaskDroid~\cite{DBLP:conf/kbse/ZhengL0ZYLC24} & 2024 & PK & \textbf{---} & \textbf{---} & Int. & Int. & \cmark & \xmark & \defenseicon	& \apkicon	\\
        & AutoRobust~\cite{DBLP:conf/raid/TsingenopoulosC24} & 2024 & PK & \textbf{---} & \textbf{---} & Int. & Int. & \cmark & \cmark & \defenseicon	& \exeicon	\\
        
		\midrule
  
        \multirow{25}{*}{\begin{tabular}[c]{@{}c@{}} \MakeUppercase{\romannumeral 4} \end{tabular}} & Mimicus~\cite{DBLP:conf/sp/SrndicL14} & 2014  & LK & \textbf{---} & \textbf{---} & \textbf{---}	& Int. & \cmark	& \xmark 	& \attackicon	& \pdficon	\\
        & EvadeML~\cite{DBLP:conf/ndss/XuQE16} & 2016  & ZK & \textbf{---} & \textbf{---} & \textbf{---} & Int. & \cmark	& \xmark	& \attackicon	& \pdficon	\\
        & EvadeHC~\cite{DBLP:conf/ccs/DangHC17} & 2017  & ZK & \textbf{---} & \textbf{---} & \textbf{---} & Int. & \xmark	& \xmark & \attackicon	& \pdficon	\\
        & Grosse \textit{et al}.~\cite{DBLP:conf/esorics/GrossePMBM17} & 2017  & PK & \textbf{---} & \textbf{---} & \textbf{---} & Int. & \xmark	& \xmark 	& \attackicon	& \apkicon	\\
        & Transcend~\cite{DBLP:conf/uss/JordaneySDWPNC17} & 2017 & PK & \textbf{---} & \textbf{---} & \textbf{---} & Ava. & \cmark & \xmark & \defenseicon	& \apkicon	\\
		& Pierazzi \textit{et al}.~\cite{DBLP:conf/sp/PierazziPCC20} & 2020  & PK & \textbf{---} & \textbf{---} & \textbf{---}	& Int. & \cmark	& \xmark 	& \attackicon	& \apkicon	\\
		& Android HIV~\cite{DBLP:journals/tifs/0002LWW0N0020} & 2020  	& PK & \textbf{---} & \textbf{---} & \textbf{---} & Int. & \xmark	& \xmark & \attackicon	& \apkicon	\\
		& Rosenberg \textit{et al}.~\cite{DBLP:conf/acsac/RosenbergSER20} & 2020  & LK & \textbf{---} & \textbf{---} & \textbf{---} & Int. 	& \xmark & \xmark	& \attackicon	& \exeicon	\\
		& HRAT~\cite{DBLP:conf/ccs/ZhaoZZZZLYYL21} & 2021  & LK & \textbf{---} & \textbf{---} & \textbf{---} & Int. & \cmark	& \xmark 	& \attackicon	& \apkicon	\\
		& OFEI~\cite{DBLP:journals/corr/abs-2105-11593} & 2021 	& LK & \textbf{---} & \textbf{---} & \textbf{---} & Int. & \xmark	& \xmark	& \attackicon	& \apkicon	\\
		& GAMMA~\cite{DBLP:journals/tifs/DemetrioBLRA21} & 2021  	& ZK & \textbf{---} & \textbf{---} & \textbf{---} & Int. & \xmark & \cmark & \attackicon	& \exeicon	\\
		& Lucas \textit{et al}.~\cite{DBLP:conf/asiaccs/LucasSBRS21} & 2021 & LK & \textbf{---} & \textbf{---} & \textbf{---} & Int. & \xmark	& \cmark & \attackicon	& \exeicon	\\
        & Li \textit{et al}.~\cite{DBLP:journals/compsec/LiL21} & 2021 & ZK & \textbf{---} & \textbf{---} & \textbf{---} & Int.	&  \xmark	& \xmark	& \attackicon	& \exeicon	\\
        & CADE~\cite{DBLP:conf/uss/Yang0HCAX021} & 2021  & PK & \textbf{---} & \textbf{---} & \textbf{---} & Ava. & \cmark	& \xmark & \defenseicon	& \apkicon	\\
        & Li \textit{et al}.~\cite{DBLP:conf/acsac/LiQCLX21} & 2021  & PK & \textbf{---} & \textbf{---} & \textbf{---} & Int. &  \xmark	& \xmark	& \defenseicon	& \apkicon	\\	
	  & MAB~\cite{DBLP:conf/asiaccs/SongLAGKY22} & 2022 	& ZK & \textbf{---} & \textbf{---} & \textbf{---} & Int. & \cmark	& \cmark	& \attackicon	& \exeicon	\\
		& BagAmmo~\cite{DBLP:conf/uss/0008CWYGYL23} & 2023  & LK & \textbf{---} & \textbf{---} & \textbf{---} & Int. & \xmark & \cmark & \attackicon	& \apkicon	\\
		& Zhang \textit{et al}.~\cite{DBLP:journals/tdsc/ZhangLCC23} & 2023 & LK & \textbf{---} & \textbf{---} & \textbf{---} & Int. & \xmark & \xmark  & \attackicon	& \exeicon	\\
		& URET~\cite{DBLP:conf/uss/EykholtLSJMZ23} & 2023 & LK & \textbf{---} & \textbf{---} & \textbf{---}	& Int. & \cmark & \xmark  & \attackicon	& \exeicon	\\
		& AdvDroidZero~\cite{DBLP:conf/ccs/HeX0J23} & 2023  & ZK & \textbf{---} & \textbf{---} & \textbf{---} & Int. & \cmark & \cmark 	& \attackicon	& \apkicon	\\
        & Sun \textit{et al}.~\cite{DBLP:conf/sigsoft/SunXTDLWZCN23} & 2023  & PK & \textbf{---} & \textbf{---} & \textbf{---} & Int. & \cmark & \cmark  	& \attackicon	& \apkicon \exeicon	\\
        & Rigaki \textit{et al}.~\cite{DBLP:journals/compsec/RigakiG23} & 2023  & LK & \textbf{---} & \textbf{---} & Con. & \textbf{---} & \xmark & \xmark & \attackicon	& \apkicon	\\
		& TRANSCENDENT~\cite{DBLP:conf/sp/BarberoPPC22} & 2023  	& PK & \textbf{---} & \textbf{---} & \textbf{---} & Ava. & \cmark & \xmark  & \defenseicon	& \apkicon \exeicon \pdficon	\\
        & MalProtect~\cite{DBLP:journals/tifs/RashidS23} & 2023 & PK & \textbf{---} & \textbf{---} & \textbf{---}	& Int. & \xmark & \xmark  & \defenseicon & \apkicon \exeicon	\\
        & MalGuise~\cite{DBLP:conf/uss/0001W0DWJLW24} & 2024 & ZK & \textbf{---} & \textbf{---} & \textbf{---}	& Int. & \cmark & \cmark  & \attackicon &  \exeicon	\\
        
\end{longtable}

\subsection{Security Risk Assumptions}
\label{sec:assumptions}

The threat model also contains knowledge and capabilities, according to previous works~\cite{DBLP:journals/corr/abs-2212-14315,DBLP:journals/corr/abs-1902-06705,DBLP:conf/sp/PierazziPCC20,DBLP:conf/ccs/HeX0J23}.
We utilize the knowledge and capabilities to represent the assumption of the security risks.

\paragraphbe{Knowledge}
Drawing from an ML-based MD system $\mathbb{F} = \{\mathbf{D}, \mathcal{F}, \mathcal{C}\} $ described in Section~\ref{sec:mlmds}, the knowledge can be categorized into perfect knowledge, limited knowledge, and zero knowledge.

In the perfect knowledge setting, the adversary has all the knowledge of the three components, which means that the adversary fully accesses the software dataset, feature space, and classifier.
In the limited knowledge setting, the adversary can only access part of the three components.
For instance, the adversary may be aware of the feature space but remains oblivious to the training dataset and the ML classifier.
In the zero knowledge setting, the adversary remains uninformed about all three components.
However, the adversary can acquire certain open-source knowledge, such as access to publicly available benign software from software markets or datasets.
In practice, the adversary usually has zero knowledge about the target ML-based MD system.

In contrast, the defender is usually the developer of ML-based MD systems.
Hence, the defender has the perfect knowledge of the three components of ML-based MD systems.
However, as attack strategies continue to evolve, defenders might be abreast of known tactics but could remain uninformed about emerging threats.
Additionally, dedicated defenders may be aware of a small group of samples introduced by the adversary.

\paragraphbe{Capabilities}
It is generally assumed that adversaries can manipulate the software samples.
In some contexts, they might even have the capability to poison the training dataset of the target ML-based MD system.
This assumption is based on the fact that some real-world antivirus solutions (e.g., VirusTotal~\cite{VirusTotal}) involve human votes to label the software samples.
In practice, the adversary usually has a limited ability to query the target model as more queries imply higher financial costs and an increased probability of detection.
Moreover, the feedback of the target ML-based MD system is available to the attacker in some cases, e.g., VirusTotal provides the detection results of every antivirus engine.
However, there might also be some cases in which the attacker cannot easily inspect the output of the target ML-based MD system~\cite{DBLP:journals/corr/abs-2212-14315}.

In contrast, the defender can conduct extensive analysis, including static analysis and dynamic analysis of the samples.
Being typically involved in the development phase of ML-based MD systems, defenders can allocate significant computational resources during this phase, even if it entails ramping up training computational demands.
However, it is vital for defenders to minimize resource-intensive computations during deployment.
This is because the high computational operations may influence the detection efficacy of the ML-based MD systems, which probably influences the user experience.

\subsection{Decomposing Attack \& Defense Strategies}
\label{sec:strategy}

We now briefly summarize the general strategies used by existing attack and defense proposals at each stage.
Then, we provide a more holistic view of the attack and defense strategies under our stage-based taxonomy.

\paragraphbe{Stage 1}
In the building dataset stage, raw software samples are processed by $ \mathcal{P} $ to construct the dataset $ \mathbf{D} $, which is essential for developing $ \mathbb{F} $. These selected software samples can be used for training the ML classifier or evaluating its performance. Attack proposals targeting this stage focus on injecting manipulated software samples into the training dataset. Although there are no existing defense proposals at this stage, the theoretical objective would be to purify the dataset to maintain the system's integrity and availability without compromising utility.

\paragraphbe{Stage 2}
In the extracting features stage, $ \mathbb{F} $ employs program analysis to obtain the program semantics and behavior of the software sample. Following program analysis, $ \mathcal{F} $ is applied to extract software features for further processing. Attacks at this stage typically employ obfuscation techniques to prevent program analysis from accurately identifying malware features. In response, defense proposals aim to extract robust feature representations through feature selection, minimizing irrelevant and potentially harmful features.

\paragraphbe{Stage 3}
In the creating an ML model stage, $ \mathbb{F} $ employs ML algorithms (e.g., SVM) to construct $ \mathcal{C} $.
This stage, primarily under the developer's control (typically the defender), focuses on training $ \mathcal{C} $ for $ \mathbb{F} $.
Therefore, the defense proposals at this stage aim to enhance the robustness of the ML model by designing robust model architectures and training methods.
Conversely, while there are no existing attack proposals, the theoretical objective would be to disrupt the training process, leading to crashes in training or reduced performance of the trained model.

\paragraphbe{Stage 4}
In the deploying an ML model stage, $ \mathbb{F} $ deploys $ \mathcal{C} $ in specific scenarios, such as on user devices, to classify unknown software samples.
Attack proposals at this stage utilize adversarial malware example generation methods to perturb malware samples or craft specific queries to extract the model. On the defense side, current efforts focus on preserving the integrity of detection results through adversarial malware example detection and maintaining the availability of detection results through concept drift detection.

We remark that we only summarize the existing strategies at each stage above. However, under our stage-based taxonomy, every CIA principle can be targeted at each stage. Thus, we further provide a more holistic view of the attack and defense strategies under our stage-based taxonomy in Table \ref{tab:cia_risks}.
For the dataset, leaking dataset (e.g., inference attack) breaks confidentiality, poisoning the dataset to inject a backdoor or degrade system performance, breaks integrity, and corrupts the dataset to forbid up-to-date malware dataset breaks availability.
For the feature extractor, leaking extraction function breaks confidentiality, obfuscating features to hide malicious features breaks integrity, and blocking analysis to prevent feature extraction breaks availability.
For the classifier, leaking model architecture and parameters breaks confidentiality, hijacking model (e.g., rowhammer attack~\cite{DBLP:conf/ccs/VeenFLGMVBRG16}) breaks integrity, and we are not aware of any availability risks other than traditional attacks such as DDoS.
For the detection results, a high false negative rate reduces recall, thus breaking integrity, and a high false positive rate raises false alarm, which prevents practical use of the system, thus breaking availability.
There is no confidentiality risks at the detection results stage, as the detection results are usually shared with the user.
Due to the stage-based taxonomy, this discussion serves as a guideline for developers to understand the potential risks at each stage and design corresponding defenses against existing and future attacks.

\subsection{Existing Attack \& Defense Proposals}
\label{sec:proposal_overview}

\subsubsection{Paper Review Methodology}
We start the paper review methodology by identifying seed papers published at top security computer security venues (i.e., IEEE S\&P, ACM CCS, USENIX Security, NDSS, TDSC, and TIFS) from 2019 to 2024, and lightweight snowballing~\cite{keele2007guidelines} the reference lists.
Then, we adopt the paper selection criteria to select the relevant and qualified studies in our preliminary paper list followed by Liu \textit{et al.}~\cite{DBLP:journals/csur/LiuTLL23}.
We employ the same first 5 exclusion criteria of paper selection in Liu \textit{et al.}~\cite{DBLP:journals/csur/LiuTLL23} together with the exclusion criterion which papers are not related to the security risks of ML-based MD system.

\subsubsection{Representative Proposal Overview}
We now present a comprehensive taxonomy of existing attack and defense proposals for ML-based MD systems with the stage-based taxonomy established in Section~\ref{sec:scope} and \ref{sec:strategy}.
As depicted in Table~\ref{tab:representativeworks}, we categorize these proposals by target stage, year of release, and proposal properties.
The proposal properties are further broken down by knowledge assumption, CIA principles in the components and detection results, open-source status, real-world evaluation, attack or defense classification, and malware format. We summarize 33 representative attack proposals and 26 representative defense proposals, totaling 59 proposals in Table~\ref{tab:representativeworks}.

Research concerning each stage of ML-based MD systems focuses on specific targets of the CIA principles, as well as showing a preference towards attacks or defenses in some stages.
For example, Stage 1 attacks compromise the integrity of the file dataset and detection results through dataset poisoning techniques, while in other domains such as text, dataset poisoning is found effective to compromise confidentiality as well \cite{DBLP:journals/corr/abs-2404-01231}.
In contrast, Stage 3 only has defense proposals aimed at enhancing the integrity of the ML model and detection results using robust architectures and training methods, while in other domains, Stage 3 attacks are also found effective \cite{DBLP:conf/uss/LvYLYZM023}.
Therefore, developers should prioritize specific types of attacks and defenses at each stage but also be aware of potential future risks of other types.

While attack and defense proposals across stages share the goal of influencing the integrity of detection results, they differ in stages.
For instance, Stage 1 attacks achieve this through dataset poisoning, while Stage 4 attacks generate adversarial malware examples.
Similarly, Stage 2 defenses preserve integrity through feature transformation, while Stage 3 defenses focus on robust model training.
Therefore, ML-based MD systems may face attacks at multiple stages simultaneously and can benefit from combined defenses at multiple stages.

Most attack and defense proposals target the integrity of detection results, as misclassifying malware as benign directly impacts system performance.
Recent works also consider the confidentiality risk of the ML model and the availability risk of the detection results.
Attack proposals in different stages require varying knowledge assumptions: Stage 1 attacks requires at least partial knowledge of the system and Stage 2 attacks usually requires zero knowledge.
In addition, while certain attacks are evaluated against real-world systems, none of the defense proposals have undergone real-world evaluation, leaving their effectiveness uncertain.

Finally, we remark that many proposals are not open-sourced, which hinders the reproducibility and evaluation of their effectiveness.
This is particularly harmful as Table \ref{tab:representativeworks} has shown increased interest in this area recently.
We encourage future works to release their code and datasets to facilitate the reproducibility and evaluation of their proposals and ask the community to take action to address this issue.

\section{Security Risks in Building Datasets}
\label{sec:sr1}

\subsection{Technical Progress}

\paragraphbe{Dataset Poisoning Attacks}
Stage 1 attacks primarily target the integrity of the file dataset $ \mathbf{D} $ and the detection results $o$, instantiated via dataset poisoning.
Such dataset poisoning attacks typically manipulate the raw data by injecting malicious samples or perturbing existing samples to compromise the classifier $ \mathcal{C} $.
The most common objective for existing dataset poisoning attacks is to inject malicious behaviors into the classifier, thereby causing misclassification when a specific trigger is present in the input, with the focus of misclassifying malware as benign.
Such attacks are also widely known as \emph{backdoor attacks} and these triggers are called \emph{backdoor triggers}.
Unlike backdoor attacks in other domains such as images, backdoor triggers in malware detection are often generated using specialized algorithms~\cite{DBLP:conf/uss/SeveriMCO21,DBLP:journals/asc/ZhangFLLY23,DBLP:journals/tdsc/LiCWWACX22,DBLP:conf/sp/YangCCPTPCW23,DBLP:conf/uss/TianQG0KZ23} or ML model explanation methods~\cite{DBLP:conf/ccs/GanM0JPHY022,DBLP:conf/sp/MinkBYCAVW23,DBLP:conf/ccs/GuoMXSWX18,DBLP:conf/nips/LundbergL17}.

Severi \textit{et al}.~\cite{DBLP:conf/uss/SeveriMCO21} propose a backdoor attack method that utilize the ML model explanation method to generate the trigger to attack MD systems using categorical features.
In this attack, they utilize the SHAP value~\cite{DBLP:conf/nips/LundbergL17} to identify the most vulnerable feature and propose a greedy method for trigger computation within the training dataset for the target model, indicating a perfect knowledge setting.
They also evaluate their method in the transferability attack scenarios and the experimental results show that they achieve a large accuracy drop on watermarked malware within a small poison rate.

SBV attack~\cite{DBLP:conf/uss/TianQG0KZ23} is a clean label backdoor attack that exploits the sparsity of feature subspaces of malware samples.
This attack proposes a dissimilarity metric-based candidate selection method that utilize nonconformity measures with P-values and a variation ratio-based trigger construction strategy.
The SBV attack is evaluated under both data-agnostic and data-and-model-agnostic scenarios and the experimental results show that the proposed attack can reduce the detection accuracy on watermarked malware to nearly 0\%, even when poisoning as little as 0.01\% of the benign class.

Li \textit{et al}.~\cite{DBLP:journals/tdsc/LiCWWACX22} present a stealthy backdoor attack that uses a Genetic Algorithm-based trigger selection to identify minimal and semantically plausible feature modifications that can be injected into apps.
To achieve stealthiness, this method proposes a label-reserving method to ensure the the poisoned samples are mislabeled as benign.
The experimental results show that the method achieves 99\% evasion rate with a poisoning rate as low as 0.3\%, and minimal impact on clean sample classification.

JP attack~\cite{DBLP:conf/sp/YangCCPTPCW23} introduces a selective clean-label backdoor attack designed to increase the stealthiness of backdoors against ML-based MD systems.
Unlike prior universal backdoors that aim to misclassify all triggered malware, JP specifically targets a selected malware family or set controlled by the attacker, leaving other malware and benign samples unaffected.
JP makes the trigger more stealthy by limiting its effective domain to a specific target family of malware samples using alternate optimization~\cite{DBLP:conf/ccs/PangSZJVLLW20}.
Experimental results on a large Android malware dataset show that JP achieves high attack success rates (e.g., over 90\% for many families) on the attacker's malware with negligible false positives and no degradation to main-task performance.

Except for the backdoor attacks, dataset poisoning can also degrade the performance of the ML-based MD system by manipulating the decision boundary of the classifier \cite{DBLP:conf/ccs/BiggioRAWCGR14}.
We remark that while recent works on dataset poisoning in the malware domain have focused on backdoor attacks, degrading the performance of the ML-based MD system via dataset poisoning worths further investigation, since works towards this direction are outdated and underrepresented while the potential risks are significant as well.

\paragraphbe{Dataset Poisoning Defenses}
As discussed in Section~\ref{sec:motivation}, eliminating the security risks as soon as possible is crucial.
However, our analysis reveals a significant gap in defense proposals compared to the attack proposals at Stage 1, as there is no defense about dataset processing to protect ML-based MD systems.
In principle, potential Stage 1 defenses aim to purify the dataset, thus removing samples that lead to malicious behaviors or degrade the performance of the system from the training set.
Nevertheless, defenses in later stages can also help to defend attacks in earlier stages.
It is particularly notable that existing defenses against dataset poisoning attacks at Stage 1 belong to Stage 3 and Stage 4, although they are mostly taken from other domains and evaluated in attack proposals.
For example, Severi \textit{et al}.~\cite{DBLP:conf/uss/SeveriMCO21} discuss activation clustering~\cite{DBLP:journals/corr/abs-1811-03728} and isolation forest~\cite{DBLP:conf/icdm/LiuTZ08}, and Yang \textit{et al}.~\cite{DBLP:conf/sp/YangCCPTPCW23} evaluate STRIP~\cite{DBLP:conf/acsac/GaoXW0RN19}, Neural Cleanse~\cite{DBLP:conf/sp/WangYSLVZZ19}, and MNTD~\cite{DBLP:conf/sp/XuWLBGL21}.
Additionally, Tian \textit{et al}.~\cite{DBLP:conf/uss/TianQG0KZ23} evaluate 10 different defense strategies against backdoor attacks originally developed for other domains, but find that these methods are not satisfying in the malware domain.
Later, Qi \textit{et al}.~\cite{DBLP:conf/uss/QiXWWMM23} propose a proactive machine learning approach for detecting backdoor poison samples in datasets.
They propose confusion training which trains a model on both the potentially poisoned dataset and a small set of clean samples with randomized labels.
This process breaks normal feature-label associations for clean data, so only poison samples with backdoor triggers remain memorized by the model.
Experiments results demonstrate the defense effectiveness of cofusing training against attacks proposed by Severi \textit{et al}.~\cite{DBLP:conf/uss/SeveriMCO21} in the Windows PE malware domain.

\subsection{Discussion \& Open Problems}

Regarding data poisoning attacks, the current focus is on injecting backdoors, but the degradation of the detection performance is also a significant threat to the availability of ML-based MD systems.
However, such attacks are underrepresented, with only one early work (published in 2014) discussing such threats.
Therefore, the community could discuss such risks more modernly since ML-based MD systems have evolved significantly since then.
In addition, the backdoor attacks in the malware domain are particularly effective, with extremely low poisoning rates, indicating a high stealthiness.
However, such strong attack performance is established on an impractical assumption that the dataset is randomly split.
In practice, the malware samples have the timestamps in the malware dataset.
Thus, recent ML-based MD systems split the dataset according to the timestamps~\cite{DBLP:conf/uss/PendleburyPJKC19,DBLP:conf/uss/ArpQPWPWCR22}.
This temporal nature may affect the effectiveness of the backdoor attacks, and the community should investigate this aspect further.

Regarding defense against dataset poisoning attacks, one remarkable observation is that existing defenses all belong to later stages, i.e., there is no defense about dataset processing to protect ML-based MD systems at Stage 1.
According to the discussion in Section \ref{sec:strategy}, the earlier the security risks are eliminated, the better.
Therefore, the community could investigate potential defense strategies at Stage 1, such as dataset purification and dataset selection, to protect ML-based MD systems.
Further, according to the empirical evidence, existing defenses against backdoor attacks are ineffective in the malware domain.
This indicates a need for further adaptation and development of backdoor defense methods for malware detection tasks.

\section{Security Risks in Extracting Features}

\subsection{Technical Progress}

\paragraphbe{Feature Extraction Attacks}
ML-based MD systems heavily depend on program analysis (static and/or dynamic) to extract features from software, thus any disruption in this extraction process poses a critical threat to the entire system.
There are two main avenues to attack feature extraction, namely \emph{program obfuscation} and \emph{variants generation}.
The former aims to conceal malicious features, while the latter aims to rearrange malicious and benign features such that the extracted features of the malware variant have the same distribution as a benign sample.
For example, consider a music player malware with the malicious behavior of reading locations; obfuscation aims to hide the location reading behavior, while variants generation aims to generate a variant for which reading location data seems benign, e.g., it has the same feature distribution as benign map applications.

Program obfuscation typically exploits the limit of static analysis in solving opaque constants by program control flow obfuscation \cite{DBLP:conf/acsac/MoserKK07} and incompleteness of static analysis by code transformation \cite{DBLP:journals/tifs/RastogiCJ14,DBLP:journals/compsec/MaiorcaACAG15,DBLP:conf/icse/HammadGM18,DBLP:journals/tdsc/WangWXZL22,DBLP:journals/softx/AonzoGVM20}.
Early attacks on static analysis feature extraction use traditional obfuscation strategies \cite{DBLP:journals/tifs/RastogiCJ14,DBLP:journals/compsec/MaiorcaACAG15} such as repacking, reflection, string encryption and class encryption, etc., while recent attacks \cite{DBLP:journals/tdsc/WangWXZL22} employ reinforcement learning to generate code obfuscation sequences.

Rastogi \textit{et al}.~\cite{DBLP:journals/tifs/RastogiCJ14} evaluate the adversarial robustness of anti-virus engines by applying three trivial transformations, eight transformations detectable by static analysis and two transformations non-detectable by static analysis.
The experimental results on 10 popular commercial anti-virus solutions demonstrate that malware can easily evade detection by applying trivial transformations.
Hammad \textit{et al}.~\cite{DBLP:conf/icse/HammadGM18} utilize 29 obfuscation techniques to evaluate the effectiveness of the top anti-malware products and find that the majority of anti-malware products are severely
impacted by even trivial obfuscations.

In addition, runtime packing~\cite{DBLP:journals/csur/MuralidharanCGN23,DBLP:conf/ndss/MantovaniAUMB20,DBLP:conf/sp/Ugarte-PedreroB15,DBLP:conf/ndss/AghakhaniGMLOBV20}, another obfuscation method, compresses/encrypts the software and decompresses/decrypts the code at runtime.
Mantovani \textit{et al}.~\cite{DBLP:conf/ndss/MantovaniAUMB20} conduct a large scale evaluation about 50K low-entropy Windows malware.
The experimental results show that ML-based models cannot identify low-entropy packed samples using static analysis features.
Aghakhani \textit{et al}.~\cite{DBLP:conf/ndss/AghakhaniGMLOBV20} train the ML model to detect the packed malware samples.
They find that the features extracted from the packed malware are not enough for ML-based models to identify other unseen packers and resist adversarial examples.

Regarding dynamic analysis which relies on sandbox execution, the focus is on sandbox detection, which aims to identify whether the software is running in a sandbox environment, thus enabling the malware to evade detection.
Miramirkhani \textit{et al}.~\cite{DBLP:conf/sp/MiramirkhaniANP17} finds that the wear-and-tear artifacts in the dynamic execution environments can be used as features for the sandbox detection.
Kondracki \textit{et al}.~\cite{DBLP:conf/ndss/KondrackiAMN22} utilize the Android API to extract environment-related features to train the classifier to detect the Android sandbox.
The experimental results show that the ML classifier can achieve 98.54\% accuracy in distinguishing between real Android devices and well-known mobile sandboxes.

Variants generation typically exploits the incompleteness of the extracted features.
Early works~\cite{DBLP:journals/tifs/XueM0TC0Z17,DBLP:conf/ccs/MengXCN0ZC16} resemble the defined modularized attack features and evasion features using evolutionary algorithms, while recent works~\cite{DBLP:journals/tifs/SenAA18,DBLP:journals/eswa/MuraliPV23,DBLP:conf/gecco/MuraliV22} further employ code transformation methods to generate malware variants more efficiently \cite{DBLP:journals/eswa/MuraliPV23,DBLP:conf/gecco/MuraliV22,DBLP:journals/tifs/SenAA18}.
They either apply functionality-preserved assembly code transformation~\cite{DBLP:journals/eswa/MuraliPV23,DBLP:conf/gecco/MuraliV22} or insert benign code segments~\cite{DBLP:journals/tifs/SenAA18}.
In particular, Sen \textit{et al}.~\cite{DBLP:journals/tifs/SenAA18} investigate the coevolutionary arms race between mobile malware and anti-malware, introducing a fully automated framework based on genetic programming for both generating evasive malware variants and the malware detectors.

\paragraphbe{Feature Extraction Defenses} All existing defenses against feature extraction attacks are employed at Stage 2, in contrast with defenses against Stage 1 attacks.
There are two main types of defenses against program obfuscation attack, namely \emph{program deobfuscation} and \emph{robust representation}. 

Program deobfuscation tries to extract malicious behaviors from obfuscated malware.
Early works \cite{DBLP:conf/ccs/DinaburgRSL08} traces program execution with hardware virtualization extensions, while later works focus on interested system calls and their arguments identified by dependence and control flow analysis \cite{DBLP:conf/ccs/CooganLD11} or data mining \cite{DBLP:conf/ccs/KiratV15}, thus improving recall of malicious behaviors by ignoring irrelevant code.

More recent works~\cite{DBLP:conf/uss/XueZCLG17,DBLP:conf/isw/AfonsoKMOGG18} further monitor malicious behaviors at runtime in an on-device and non-invasive way.
Xue \textit{et al}.~\cite{DBLP:conf/uss/XueZCLG17} introduce Malton, an on-device, non-invasive dynamic analysis platform designed for comprehensive behavior analysis of Android malware running under the ART runtime.
Malton provides multi-layer monitoring and information flow tracking, bridging semantic gaps between Java and native code, and tracking sensitive data flows even across JNI and reflective calls.
Additionally, it integrates a path exploration engine using concolic execution to trigger conditionally executed behaviors.

To defend against runtime packing, defense proposals typically adopt unpacking mechanisms to extract the hidden code within the packed malware.
Early works extract hidden code with hybrid analysis \cite{DBLP:conf/acsac/RoyalHDEL06} or execution monitoring \cite{DBLP:conf/acsac/MartignoniCJ07}, while later works \cite{DBLP:conf/esorics/ZhangLY15,DBLP:conf/icse/XueLYWW17,DBLP:journals/tse/XueZLYWZM22} identify collection points to recover the packed code.
More recent works \cite{DBLP:conf/uss/Cheng0LZFPM21,DBLP:conf/sp/XueZLZSGZA21,DBLP:conf/uss/ChengLZ023} unpack the packed malware via hardware-assisted approaches.

API-Xray~\cite{DBLP:conf/uss/Cheng0LZFPM21} reconstructs a working import table from an unpacked malware’s process memory at its original entry point, making the sample executable and its API calls analyzable for further static/dynamic inspection.
It utilizes Intel’s Branch Trace Store and NX bit hardware features to track the real target APIs.
Extensive experiments demonstrate that API-Xray can recover import tables for complex commercial packing products.

Happer~\cite{DBLP:conf/sp/XueZLZSGZA21} unpacks the Android apps by leveraging ARM hardware tracing and hardware breakpoints to non-invasively monitor the actual runtime behaviors on real devices.
Based on the detected behaviors, Happer then adaptively selects the best strategy to extract all hidden Dex data from memory, even when packers release code just-in-time or modify runtime objects, and reconstructs valid Dex files suitable for further static analysis.
Extensive experiments with 12 commercial packers and over 24,000 Android apps show that Happer reveals 27 distinct packing behaviors and achieves much higher unpacking coverage and fidelity than existing tools.

LoopHPCs~\cite{DBLP:conf/uss/ChengLZ023} investigate the feasibility of using hardware performance counters (HPCs) for malware unpacking, addressing recent skepticism about the non-deterministic nature of HPCs for security uses.
Specifically, it utilizes the Precise Event-Based Sampling and Last Branch Record offered by Intel CPUs to unpack malware.
Extensive experiments show that LoopHPCs achieves high unpacking accuracy and outperforms state-of-the-art software-based unpackers.

Instead of extracting obfuscated malicious behaviors in a post-hoc manner, robust representation aims to design a feature space that can hardly be obfuscated.
One way to achieve this is by transforming features into a more robust space.
Early works use feature selection strategies based on manipulation cost \cite{DBLP:conf/acsac/ChenHY17} and transform the original binary features into continuous probabilities \cite{DBLP:conf/asunam/ChenHYX18}, while more recent works capture robust API features against Android system updates \cite{DBLP:conf/ccs/ZhangZZDCZZY20} and transform the feature space into a more robust representation using an ensemble of feature transformations \cite{DBLP:journals/tifs/HanSX20}.

APIGraph~\cite{DBLP:conf/ccs/ZhangZZDCZZY20} builds a robust feature space by incorporating the evolution of the Android APIs.
It clusters the features with semantic similarity information among Android APIs by constructing an API relation graph from official Android documentation.
The experimental results demonstrate that APIGraph’s clusters capture semantic similarity within malware families and that semantically-close APIs are grouped together beyond package-level abstraction.

Han \textit{et al}.~\cite{DBLP:journals/tifs/HanSX20} propose FARM that uses irreversible feature transformations to defend against evasion attacks targeting machine learning classifiers.
FARM begins by extracting standard static, dynamic, and API package call features from Android apps, then applies three novel classes of feature transformations to map these features into a new space.

Another way is to use obfuscation-resilient representation directly.
Early works \cite{DBLP:conf/securecomm/AaferDY13,DBLP:journals/tosem/GarciaHM18,DBLP:conf/acsac/MachiryRGFCKV18} use reflection-based features or features that are prohibitively difficult to manipulate, while more recent works \cite{DBLP:conf/cikm/HouFZYLWWXS19,DBLP:conf/infocom/LingWDQZZMWWJ22,DBLP:journals/tifs/GaoCYHLYL23} use graph-related features that are resilient to most obfuscation methods.
These feature transformations make it difficult for attackers to manipulate features in a way that will fool the classifier.


CorDroid~\cite{DBLP:journals/tifs/GaoCYHLYL23} leverages the complementarity of multiple semantic features to withstand a wide variety of bytecode-level obfuscation techniques.
It employs the Enhanced Sensitive Function Call Graph and Opcode-based Markov transition Matrix as features.
Experimental results demonstrate that it achieves defense effectiveness in detecting obfuscated Android application samples.

MaskDroid~\cite{DBLP:conf/kbse/ZhengL0ZYLC24} introduces a masking-based self-supervised learning approach within a GNN framework.
In the training process, MaskDroid is tasked with reconstructing the full graph from the remaining nodes and incorporates a proxy-based contrastive module to compress intra-class variance and sharpen the decision boundary.
Extensive experiments on a large AndroZoo-sourced dataset demonstrate that MaskDroid reduces attack success rates from 41.5\% to 32\% under the white-box attacks setting, while maintaining competitive detection effectiveness and efficiency.

\subsection{Discussion \& Open Problems}

Regarding feature extraction attacks, existing program obfuscation techniques have been shown to be effective against common feature extractors, especially in the wild, since obfuscation rarely targets specific feature extractors.
However, as obfuscation adds many instructions to the original code, some even being executed, this brings memory and computation overhead to the original code.
This problem is particularly important when program obfuscation is used for good, e.g., protecting copyrights.
The other attack direction, malware variant generation, requires prior knowledge about the malicious features, which limits its generalization to malware families.
Therefore, whether the automatic creation of malware variants is possible without human-designed attack features is an open question.

Regarding defense proposals, the SOTA deobfuscation works are limited in practice since they typically require specialized hardware and software.
For instance, Malton is based on the Valgrind framework, which can be evaded by anti-emulator techniques; LoopHPCs does not support AMD processors due to the lack of hardware tracing mechanisms like LBR.
Robust feature space methods, especially those that utilize function call graph-based features, are no longer robust because recent attack proposals have leveraged adversarial machine learning techniques to manipulate function call graph-based features, thus breaking their underlying assumption.
Therefore, better robust feature space methods that are not based on function call graphs are needed.

\section{Security Risks in Creating ML Models}
\label{sec:sr3}


\subsection{Technical Progress}

Unlike other stages, while attacks at Stage 3 (model hijacking) are possible via injecting malicious code into the training framework (malicious wheel builder, e.g., open-source project contributors) and manipulating the training hardware (malicious hardware provider, e.g., cloud platforms), these attacks are not yet specialized for ML-based MD systems.
Therefore, we focus on the defense proposals at this stage.
Instead of defending against model hijacking, Stage 3 defenses defend against various attacks at different stages, namely backdoor attacks (Stage 1) and adversarial malware attacks (Stage 4).
Backdoor defenses have been discussed in Section~\ref{sec:sr1}.
Thus, we focus on defenses against adversarial malware attacks in the following.
Such defenses can be roughly categorized into \emph{robust model architecture} and \emph{robust model training}.
In the following, we illustrate each approach in more detail.

\paragraphbe{Robust Model Architecture}
Enhancing the robustness of ML-based MD systems can be achieved through inherently robust model architectures.
This involves using ensemble learning techniques~\cite{DBLP:journals/tifs/LiL20,DBLP:journals/tnse/LiLYX21,DBLP:conf/ndss/SmutzS16,DBLP:journals/corr/abs-1712-05919} by employing voting mechanisms~\cite{DBLP:journals/tifs/LiL20,DBLP:journals/tnse/LiLYX21} or the diversity of an individual classifier to identify the evasion samples~\cite{DBLP:conf/ndss/SmutzS16}.
Additionally, the defender could also utilize another model as a complement to the vanilla model to capture the features of adversarial malware examples.

KuafuDet~\cite{DBLP:journals/compsec/ChenXFHXZL18} introduces a two-phase, self-adaptive learning framework containing offline training and online detection modules.
The online detection module employs a camouflage detector to identify and re-label suspicious false negatives for iterative retraining.
Evaluations on a large dataset show that KuafuDet achieves higher detection accuracy than Drebin and is robust under severe data imbalance.

Li \textit{et al}.~\cite{DBLP:conf/www/LiZYLGC21} utilized a variational autoencoder to measure the reconstruction error of the input samples to filter out adversarial malware examples.
The system makes a final decision by aggregating the outputs of both the VAE and MLP, declaring an app benign only if both models agree.
Extensive experiments on a large real-world dataset demonstrate that this method remains effective against both white-box and black-box adversarial attacks.

The recent SOTA work~\cite{DBLP:journals/tifs/ZhouCYCH24} maintains a dynamic pool of heterogeneous models, each trained on distinct data partitions and further diversified through adversarial training with multiple perturbation strategies, thereby reducing the exposure and transferability risks associated with single-model detectors.
To optimize detection robustness, MTDroid employs a two-stage sub-classifier selection algorithm for ensemble learning and introduces a hybrid update strategy that dynamically refreshes the model pool based on query and failure statistics.
Experimental results demonstrate that MTDroid achieves state-of-the-art performance in both detection accuracy and robustness.

Another approach~\cite{DBLP:conf/kdd/WangGZOXLG17,DBLP:journals/corr/abs-1809-06498,DBLP:journals/tdsc/DemontisMBMARCG19,DBLP:conf/wcre/LiHWHLG23} to enhance the ML classifier robustness involves designing constraints to process the feature using part of the model architecture.
Wang \textit{et al}.~\cite{DBLP:conf/kdd/WangGZOXLG17} proposed an extra layer for random feature nullification between the input and the first hidden layer in DNN to achieve the robustness by making the DNN model non-deterministic.
RFN introduces stochasticity in both training and inference by randomly nullifying input features, making the model non-deterministic and significantly hindering an adversary's ability to craft effective adversarial perturbations.

Sec-SVM~\cite{DBLP:journals/tdsc/DemontisMBMARCG19} starts from the intuition of bounding classifier sensitivity and constrains the distribution of feature weights, effectively enforcing more evenly-distributed (dense) weights via $\ell_\infty$-norm regularization.
This approach significantly increases the number of feature manipulations required for a successful evasion.
Experiments on large-scale real-world Android datasets show that Sec-SVM achieves state-of-the-art robustness across a variety of attack scenarios, outperforming both standard SVM and previously proposed bagging-based approaches.

Monotonic classifiers~\cite{DBLP:journals/corr/abs-1806-06108,DBLP:conf/codaspy/IncerTA018} present another intriguing avenue for the integrity preserving.
These classifiers maintain or increase the confidence in a malware label, irrespective of any additional features added to a sample.
Most of the current adversarial malware example attacks only add benign features to the malware samples due to the problem space constraints, which cannot bypass the defense of the monotonic classifier.

\paragraphbe{Robust Model Training}
The robustness of an ML classifier can also be achieved by utilizing robust model training methods.
Adversarial training is the most common solution to train a robust model in the image domain~\cite{DBLP:journals/corr/GoodfellowSS14}.
The key intuition of adversarial training is incorporating adversarial examples in the process of training an ML classifier.
To adapt the adversarial training techniques to the malware domain, current defense proposals~\cite{DBLP:conf/sp/Al-DujailiHHO18,DBLP:journals/prl/RathoreSSS22,DBLP:journals/corr/abs-2110-11987,DBLP:journals/corr/abs-2302-11328,DBLP:conf/securecomm/PodschwadtT19,DBLP:journals/eaai/ShaukatLV22,DBLP:conf/ausai/ChenY17,DBLP:conf/eisic/ChenYB17} propose the customized training loss and adversarial malware example attacks in the feature space for adversarial training.
The earliest works~\cite{DBLP:conf/ausai/ChenY17,DBLP:conf/eisic/ChenYB17} introduce the extra regularization term in the training loss to enhance the training robustness.
The later works~\cite{DBLP:conf/sp/Al-DujailiHHO18,DBLP:journals/corr/abs-2110-11987,DBLP:journals/prl/RathoreSSS22,DBLP:journals/eaai/ShaukatLV22} focus on generating adversarial examples for adversarial training.
For instance, Al-Dujaili \textit{et al}.~\cite{DBLP:conf/sp/Al-DujailiHHO18} proposed the Bit Gradient Ascent method designed for the discrete feature.
Galovic \textit{et al}.~\cite{DBLP:journals/corr/abs-2110-11987} proposed the adversarial strings generation method from the perturbed latent space.
The recent SOTA work~\cite{DBLP:journals/corr/abs-2302-11328} designs the Stepwise Mixture of Attacks with convergence guarantees for searching adversarial perturbations in the feature space.

To explore the effectiveness of adversarial training in the domain of raw-binary malware detectors, Lucas \textit{et al}.~\cite{DBLP:conf/uss/LucasPLBRS23} utilize adversarial training for raw-binary malware detection by generating adversarial examples through In-Place Replacement, Displacement, and Kreuk attacks.
Their experiments reveal that standard data augmentation with random, functionality-preserving binary transformations provides little to no robustness against guided adversarial attacks.
In contrast, adversarial training using guided, attack-specific adversarial examples significantly increases model robustness.

AutoRobust~\cite{DBLP:conf/raid/TsingenopoulosC24} utilizes a reinforcement learning-based adversarial training methodology to enhance the robustness of the ML-based MD systems based on dynamic analysis reports.
It formulates the generation of adversarial examples as a Markov Decision Process, where an RL agent learns functionality-preserving transformations directly on dynamic analysis reports, ensuring all modifications are realistic and feasible for actual malware binaries.

However, adversarial training offers only empirical robustness of the ML classifier and may be susceptible to stronger attacks~\cite{DBLP:conf/icml/Croce020a}.
Thus, certified training~\cite{DBLP:conf/sp/GehrMDTCV18,DBLP:conf/nips/ZhangWCHD18,DBLP:conf/iclr/MaoM0V24,DBLP:journals/corr/abs-2403-07095,DBLP:journals/corr/abs-2406-04848} has been proposed to provide certified robustness of ML classifiers in the image domain~\cite{mueller2023certified,mao2023connecting,DBLP:conf/iclr/PalmaBDKSL24} and text domain~\cite{DBLP:conf/ccs/DuJSZLSFYB021}.
Current defense proposals~\cite{DBLP:conf/ccs/GibertZL23,huang2023rs} in the malware domain also adapt certified defenses.

Chen \textit{et al}.~\cite{DBLP:conf/uss/Chen0SJ20} propose a novel subtree distance metric on the PDF parse tree and specify two main classes of robustness properties: subtree insertions and subtree deletions under the PDF root.
It leverages the verifiable robust training methods~\cite{DBLP:conf/nips/WangPWYJ18,DBLP:conf/uss/WangPWYJ18} to build neural network classifiers that are formally guaranteed to be robust against any attacker bounded by these properties.

RS-Del~\cite{huang2023rs} extends randomized smoothing~\cite{DBLP:conf/icml/CohenRK19,DBLP:conf/icml/YangDHSR020} to the edit distance setting by applying random deletions to input sequences.
It only requires random deletions to certify robustness against all three edit operations, leveraging a proof that relies on longest common subsequence analysis rather than the standard Neyman-Pearson approach.
It achieves a certified accuracy of 91\% in the MalConv model at an edit distance radius of 128 bytes, with only a marginal drop in clean accuracy.

\subsection{Discussion \& Open Problems}

Similar to other domains, increasing the adversarial robustness of ML-based MD systems via architecture design or specialized training methods often comes at the cost of clean accuracy and computational overhead.
While this is more acceptable since attacks on MD systems are more common, it is unclear whether this trade-off can be resolved.
For instance, FD-VAE~\cite{DBLP:conf/www/LiZYLGC21} requires extensive hyperparameter tuning to maintain clean accuracy at the cost of multiplied training time.
Moreover, robust training typically relies on specific train-time attacks, but the adversarial robustness is poorly generalized against other attacks.
In particular, feature-space attacks are more efficient and frequently adopted in training, but such attacks often yield mismatches with realistic problem-space attacks due to different attack specifications.
This often brings difficulty in having more practical certified robustness, as the input specification in the feature space is not norm-based, but current certified defenses only consider norms such as $L_0$ (patch) distance because such normed specifications are more established in other domains.
In addition, while defenses against backdoor and adversarial malware attacks are extensively discussed, the confidentiality of the ML model is largely overlooked since no existing defense in the malware domain protects confidentiality, e.g., differential privacy.

\section{Security Risks in Deploying ML Models}
\label{sec:sr4}


\subsection{Technical Progress}

This section is divided into four parts: model extraction attack (targeting confidentiality), adversarial malware attack (targeting integrity), adversarial malware defense (preserving integrity) and concept drift detection (preserving availability).
This is because currently no specialized defense is designed to counter the model extraction attack.
The availability risks at this stage are mainly caused by concept drift (similar to out-of-distribution in other domains), a phenomenon in which malware distribution shifts over time.
This shift leads to high false positive rates and influences the availability of the detection results.

\paragraphbe{Model Extraction}
Because detection results are shared with users in Stage 4, the ML models are exposed to model extraction attacks~\cite{DBLP:conf/uss/TramerZJRR16,DBLP:conf/uss/JagielskiCBKP20,DBLP:conf/ndss/YuYZTHJ20,DBLP:conf/uss/HorvathL0B24,DBLP:conf/uss/NayanGABUS24}.
Model extraction attacks in other domains mainly extract the model functionality, model architecture and model parameter.
However, in the malware domain, the feature extraction function $ \mathcal{F} $ is worth extraction as well.
Early works~\cite{DBLP:conf/ccs/WressneggerFYR17,DBLP:conf/codaspy/CaiY16} extract the malware signatures and the detection logic of the anti-virus engines, and more recent works~\cite{DBLP:journals/compsec/RigakiG23} focus on stealing ML model functionality.

Rigaki \textit{et al}.~\cite{DBLP:journals/compsec/RigakiG23} propose a model stealing attack to extract the model functionality of commercial antivirus engines.
They design the dualFFNN, which leverages true labels with a skip connection for stable training, and the FFNN-TL, which combines transfer learning with active learning for improved surrogate accuracy.

\paragraphbe{Adversarial Malware Attack}
To subvert the integrity of the detection results, current attacks utilize adversarial malware generation methods by perturbing the malware to evade detection.
Most of the adversarial malware generation methods~\cite{DBLP:journals/corr/GrossePM0M16,DBLP:conf/esorics/GrossePMBM17,DBLP:journals/tifs/0002LWW0N0020,DBLP:conf/ccs/ZhaoZZZZLYYL21,DBLP:conf/eusipco/KolosnjajiDBMGE18,DBLP:conf/sp/SuciuCJ19,DBLP:conf/asiaccs/LucasSBRS21,DBLP:conf/iwqos/Liu0LL19,DBLP:journals/corr/abs-1708-06131,DBLP:conf/nca/AbaidKJ17} assume the attacker has the perfect knowledge or limited knowledge of ML-based MD systems.
These assumptions are supported by the Kerckhoffs' principle~\cite{kerckhoff} and represent the worst cases of the defender~\cite{DBLP:conf/sp/PierazziPCC20}.
Under these scenarios, the adversarial malware generation algorithms may leverage optimization methods utilizing gradient-based optimization~\cite{DBLP:journals/corr/GrossePM0M16,DBLP:conf/esorics/GrossePMBM17,DBLP:journals/tifs/0002LWW0N0020,DBLP:conf/ccs/ZhaoZZZZLYYL21,DBLP:conf/sp/SrndicL14} or gradient-free optimization methods~\cite{DBLP:conf/acsac/RosenbergSER20,DBLP:journals/corr/abs-2105-11593,DBLP:conf/uss/0008CWYGYL23,DBLP:conf/codaspy/KucukY20}.
In gradient-based optimization methods, the attacker may select perturbations with the largest gradients of the objective function.
For example, Android HIV~\cite{DBLP:journals/tifs/0002LWW0N0020} refines the C\&W attack~\cite{DBLP:conf/sp/Carlini017} and JSMA attack~\cite{DBLP:conf/eurosp/PapernotMJFCS16} to generate adversarial malware examples in the feature space.
HRAT~\cite{DBLP:conf/ccs/ZhaoZZZZLYYL21} selects the manipulation type and manipulation targets using the gradient descent method.
In gradient-free optimization methods, the attacker may leverage the hill-climbing method~\cite{DBLP:conf/ccs/DangHC17}, generative adversarial networks~\cite{DBLP:conf/acsac/RosenbergSER20,DBLP:journals/sj/LiZYLL20}, and genetic algorithms~\cite{DBLP:conf/ndss/XuQE16,DBLP:conf/uss/0008CWYGYL23}.
The recent SOTA work~\cite{DBLP:conf/uss/0008CWYGYL23} designs the generative adversarial network combined with a multi-population co-evolution algorithm to generate the desired perturbation.
Moreover, the attacker~\cite{DBLP:conf/itasec/DemetrioBLRA19} could also find the most vulnerable features in the feature space directly under the perfect knowledge setting.
For instance, Pierazzi \textit{et al}.~\cite{DBLP:conf/sp/PierazziPCC20} leveraged the weights of the SVM classifier to assess feature importance and subsequently searched for the most benign features to inject.
Wang \textit{et al}.~\cite{DBLP:conf/sigsoft/SunXTDLWZCN23} employed the SHAP values~\cite{DBLP:conf/nips/LundbergL17} to compute the accrued malicious magnitude of features for manipulation.

In practical scenarios, the internals of ML-based MD systems typically remain unknown to the attackers.
Under the zero knowledge setting, existing attacks utilize extra knowledge inside the program semantics~\cite{DBLP:conf/acsac/YangK0G17,DBLP:journals/tifs/DemetrioBLRA21,DBLP:journals/tdsc/ZhangLCC23,DBLP:journals/corr/abs-2110-03301,DBLP:conf/uss/EykholtLSJMZ23,DBLP:journals/tifs/ZhanDHLGP24,DBLP:conf/raid/RosenbergSRE18,DBLP:journals/compsec/ZhongHZLC22} and query feedback~\cite{DBLP:journals/compsec/LiL21,DBLP:journals/sis/RathoreSSS22,DBLP:conf/asiaccs/SongLAGKY22,DBLP:conf/ccs/HeX0J23,DBLP:conf/uss/0001W0DWJLW24}.
They may involve mimicking benign samples~\cite{DBLP:conf/acsac/YangK0G17}, incorporating benign program slices~\cite{DBLP:journals/tifs/DemetrioBLRA21,DBLP:journals/corr/abs-2110-03301} or leveraging the malware program structure~\cite{DBLP:journals/tdsc/ZhangLCC23}.
Moreover, they could also define the malware perturbations and utilize the query feedback to adjust the selection probability of these perturbations.
The perturbation selection process can be modeled as the reinforcement learning process~\cite{DBLP:journals/corr/abs-1801-08917,DBLP:journals/compsec/LiL21,DBLP:journals/sis/RathoreSSS22,DBLP:conf/asiaccs/SongLAGKY22}.
The recent SOTA works~\cite{DBLP:conf/ccs/HeX0J23,DBLP:conf/uss/0001W0DWJLW24} utilize the perturbation selection tree, incorporating the perturbation semantics or Monte Carlo tree search for the effective perturbation selection.

He \textit{et al}.~\cite{DBLP:conf/ccs/HeX0J23} propose AdvDroidZero, generating adversarial Android malware samples under the zero knowledge setting.
AdvDroidZero designs a perturbation selection tree that clusters the semantically similar perturbations together to improve the attack effectiveness and efficiency.
The experimental results demonstrate that AdvDroidZero can achieve about 90\% attack success rate against real-world anti-virus engines.

Ling \textit{et al}.~\cite{DBLP:conf/uss/0001W0DWJLW24} propose MalGuise, a practical black-box adversarial attack framework targeting learning-based Windows malware detection systems.
MalGuise introduces a semantics-preserving transformation called call-based redividing, which manipulates both nodes and edges of a malware’s control-flow graph and leverages a Monte Carlo Tree Search algorithm to efficiently search for optimized sequences of transformations.
Experimental results demonstrate that MalGuise achieves 74.97\% against real-world anti-virus solutions.

Properties of adversarial malware attacks have gained attention as well.
Prior works~\cite{DBLP:conf/uss/DemontisMPJBONR19,DBLP:conf/sp/MaoFWJ0LZLB022} explore the transferability of adversarial examples, showing that similar conclusions hold for adversarial malware examples as well.
Demetrio \textit{et al}.~\cite{DBLP:journals/tissec/DemetrioCBLAR21} benchmark 7 adversarial malware example attacks in the context of Windows PE malware.
They suggest that the size of injected bytes correlates with stealthiness, and model architecture has a greater impact on model robustness than training dataset size and activation function.
Recent works \cite{DBLP:journals/corr/abs-2102-06747,DBLP:journals/tifs/ZhanDHLGP24} generate universal adversarial perturbations and patches to understand perturbation adaptability across malware samples.

\paragraphbe{Adversarial Malware Defense}
Defenses against adversarial malware examples either try to design robust models at Stage 3 (discussed in Section \ref{sec:sr3}) or purify and abstain from adversarial malware examples at Stage 4.
Rashid \textit{et al}.~\cite{DBLP:journals/tifs/RashidS23} present MalProtect, which employs a suite of diverse threat indicators, including statistical and autoencoder-based anomaly scores, to analyze sequences of user queries before they reach the prediction model.
These indicator scores are aggregated by a learned decision model to detect ongoing attacks.
Experimental results across Android and Windows malware datasets demonstrate that MalProtect can reduce the evasion rate of query attacks by over 80\%.
To purify queries, Zhou \textit{et al}.~\cite{DBLP:journals/corr/abs-2312-06423} employ a denoising autoencoder framework to reconstruct the input, thus removing adversarial perturbations.

\paragraphbe{Concept Drift Detection}
Concept drift is a phenomenon where the dynamics of malware in the wild lead to performance degradation of the ML-based MD system.
Hence, the current defense proposals~\cite{DBLP:conf/uss/JordaneySDWPNC17,DBLP:conf/uss/Yang0HCAX021,DBLP:conf/sp/BarberoPPC22} address this problem by detecting the concept drift samples to maintain the detection effectiveness.

Transcend~\cite{DBLP:conf/uss/JordaneySDWPNC17} employs a conformal evaluator to compute non-conformity measures and p-values for incoming samples, assessing the statistical fit of each prediction relative to the training data.
By monitoring algorithm credibility and confidence, Transcend can flag untrustworthy predictions and identify drifting samples in both binary and multi-class settings.
TRANSCENDENT~\cite{DBLP:conf/sp/BarberoPPC22} builds upon this, refining the conformal evaluation theory and improving the efficiency of Transcend by calibration splits.
To explain the concept drift samples, CADE~\cite{DBLP:conf/uss/Yang0HCAX021} leverages autoencoder with contrastive learning to project the sample into the monitoring space and uses distance changes to explain the set of important features.

Another approach to address concept drift is the fine-tuning of the ML model.
Chen \textit{et al}.~\cite{DBLP:conf/uss/0001D023} propose a novel continuous learning framework that combines hierarchical contrastive learning and a new pseudo loss uncertainty measure to guide active learning and sample selection. 
The hierarchical contrastive learning explicitly models the similarity among malware families and between benign/malicious apps, improving generalization and detection of new malware families under severe class imbalance.
Experimental results on the dataset spanning 7+ years show that this approach reduces the false negative rate from 14\% to 9\% and the false positive rate from 0.86\% to 0.48\%.

\subsection{Discussion \& Open Problems}

Stage 4 attacks are the most extensively discussed, potentially due to the direct interaction between the ML-based MD system and the users.
Current model extraction attacks are constrained by high query costs in order to be effective.
In addition, the literature only discusses extraction attacks for certain malware types, which makes its generalization to other malware types questionable.
Such extraction attacks also make strong knowledge assumptions, e.g., the knowledge about the model architecture.
Future research could explore hardware-based model extraction attacks, such as side-channel reverse engineering, to reduce the cost and improve the effectiveness of these attacks.
Among all the possible attacks, adversarial malware examples receive the most attention.
However, SOTA adversarial malware attacks often disregard stealthiness, as they insert unlimited pieces of benign codes for evasion.
Such attacks can hardly bypass monotonic classifiers in principle, because the inserted benign codes cannot reduce the maliciousness of the malware.
Future research could discuss enhancing the stealthiness of adversarial malware examples by eliminating existing malware features without impacting the malicious functionality.

Arguably, the deployment stage is the most vulnerable to attacks, where the ML-based MD system is directly exposed to the users.
However, the literature on defense at Stage 4 is limited, while defenses against attacks at this stage focus on Stage 3.
In particular, no defense can identify adversarial malware examples based on individual queries, with few defenses relying on monitoring the continuous query history and purifying the input.
Such defenses may be easily evaded if the attacker is aware of the defense mechanism.
In addition, methods used to detect concept drifts are similar to those used to detect adversarial malware examples.
Future research could explore the possibility of unifying and generalizing these two detection methods.

\section{Delving into Stage-wise Interconnections} \label{sec:casestudy}

We present a stage-based taxonomy, which decomposes the security risks in ML-based MD systems into operational stages.
In this section, we will focus on the interconnections between these stages to provide new empirical insights.
Since the taxonomy is designed to be comprehensive, such insights cannot be exhaustive.
Thus, we conduct two specific case studies: one inter-stage study connecting poisoning and evasion attacks (Stages 1 and Stage 4) and one intra-stage study connecting evasion and concept drift detection (Stage 4).
These case studies aim to provide a deeper understanding of the interconnections between different stages and to inspire future research directions.

\paragraphbe{Dataset}
Our dataset comprises 134,759 benign applications and 14,775 malware samples, totaling 149,534 applications between January 2015 and October 2016.
This dataset is the same as the previous work~\cite{DBLP:conf/sp/YangCCPTPCW23}.
We choose this dataset because the malware family information is directly available, which is required by the JP attack.

\paragraphbe{Implementation Details}
To provide case studies, we have developed an evaluation framework grounded in the stage-based taxonomy approach.
This framework implements the ML-based MD methods following the stage-based design, integrating the SOTA attack and defense proposals~\cite{DBLP:conf/ccs/HeX0J23,DBLP:conf/sp/BarberoPPC22,DBLP:conf/sp/YangCCPTPCW23}.
Our case studies focus on the Android malware domain, a choice motivated by the abundance of data resources available in this area.
For example, AndroZoo~\cite{DBLP:conf/msr/AllixBKT16} has over 24 million APKs with rich analysis results.
Besides, the current SOTA proposals~\cite{DBLP:conf/ccs/HeX0J23,DBLP:conf/sp/BarberoPPC22,DBLP:conf/sp/YangCCPTPCW23} primarily focus on the Android domain.
However, our evaluation framework can be easily extended into other malware formats due to the stage-based design.
Other potential users only need to replace the software feature analysis module with the corresponding feature extractor, which requires relatively small workloads.

\begin{table}[t]
    \setlength{\abovecaptionskip}{0pt}
	\caption{The attack performance of adversarial example attacks measured by attack success rate under poisoned and benign models. ``Feature, Problem, Non Problem'' represent the trigger constraints.}
 
	\begin{tabular}[centering,width=0.5*\linewidth]{@{}C{2.0cm}C{1.5cm}C{1.5cm}C{1.5cm}C{1.5cm}C{1.5cm}C{1.5cm}@{}}
		\toprule
        \multirow{3}{*}{Family} & \multicolumn{2}{c}{Feature} & \multicolumn{2}{c}{Problem} & \multicolumn{2}{c}{Non Problem} \\ \cmidrule(l){2-3} \cmidrule(l){4-5} \cmidrule(l){6-7}
         & Poison Model & Benign Model & Poison Model & Benign Model & Poison Model & Benign Model \\
        \midrule
        mobisec & 0.38 & 0.28 & 0.40 & 0.28 & 0.23 & 0.29 \\
        tencentprotect & 0.33 & 0.22 & 0.41 & 0.34 & 0.41 & 0.20 \\
        anydown & 0.46 & 0.36 & 0.37 & 0.33 & 0.23 & 0.30 \\
        leadbolt & 0.43 & 0.31 & 0.47 & 0.31 & 0.37 & 0.29 \\
        eldorado & 0.28 & 0.29 & 0.18 & 0.32 & 0.34 & 0.34 \\
        jiagu & 0.33 & 0.29 & 0.29 & 0.32 & 0.30 & 0.22 \\
        kuguo & 0.13 & 0.31 & 0.31 & 0.31 & 0.21 & 0.32 \\
        \bottomrule
  	\end{tabular}
 
	\label{tab:AdvPoisonRes}
\end{table}

\begin{table}[t]
    \setlength{\abovecaptionskip}{0pt}
	\caption{The attack results of the JP attack under different trigger spaces and different target families. ``AT, AR, FB, F1'' = Attack Success Rate in samples with target family, Attack Success Rate in samples without target family, False Positive Rate in benign samples, F1 scores in malware detection task.}
 
	\begin{tabular}[centering,width=0.5*\linewidth]{@{}C{2.0cm}C{1.5cm}C{2.0cm}C{1.6cm}C{0.85cm}C{0.85cm}C{0.85cm}C{0.85cm}@{}}
		\toprule
        Target Family & Apps Nums & Space & Trigger Size & AT & AR & FB & F1 \\
		\midrule
        \multirow{3}{*}{mobisec} & \multirow{3}{*}{57} & Feature & 22 & 1.00 & 0.14 & 0.00 & 0.90 \\
         &  & Problem & 14 & 0.64 & 0.02 & 0.04 & 0.92 \\
         &  & Non-problem & 25 & 1.00 & 0.35 & 0.07 & 0.92 \\
        \midrule
        \multirow{3}{*}{tencentprotect} & \multirow{3}{*}{157} & Feature & 46 & 0.98 & 0.51 & 0.00 & 0.92 \\
         &  & Problem & 37 & 0.98 & 0.42 & 0.00 & 0.92 \\
         &  & Non-problem & 16 & 0.74 & 0.23 & 0.00 & 0.91 \\
        \midrule
        \multirow{3}{*}{anydown} & \multirow{3}{*}{192} & Feature & 20 & 0.95 & 0.03 & 0.02 & 0.91 \\
         &  & Problem & 19 & 1.00 & 0.24 & 0.00 & 0.92 \\
         &  & Non-problem & 18 & 0.98 & 0.09 & 0.00 & 0.92 \\
        \midrule
        \multirow{3}{*}{leadbolt} & \multirow{3}{*}{222} & Feature & 34 & 0.91 & 0.13 & 0.00 & 0.91 \\
         &  & Problem & 28 & 0.56 & 0.06 & 0.00 & 0.92 \\
         &  & Non-problem & 24 & 0.93 & 0.68 & 0.00 & 0.91 \\
        \midrule
        \multirow{3}{*}{eldorado} & \multirow{3}{*}{335} & Feature & 32 & 0.88 & 0.32 & 0.00 & 0.92 \\
         &  & Problem & 51 & 0.92 & 0.37 & 0.00 & 0.91 \\
         &  & Non-problem & 26 & 0.72 & 0.61 & 0.00 & 0.92 \\
        \midrule
        \multirow{3}{*}{jiagu} & \multirow{3}{*}{668} & Feature & 38 & 0.84 & 0.16 & 0.00 & 0.91 \\
        &  & Problem & 42 & 0.96 & 0.48 & 0.00 & 0.92 \\
        &  & Non-problem & 22 & 0.90 & 0.80 & 0.00 & 0.92 \\
        \midrule
        \multirow{3}{*}{kuguo} & \multirow{3}{*}{2,897} & Feature & 28 & 0.90 & 0.26 & 0.00 & 0.92 \\
         &  & Problem & 37 & 0.83 & 0.33 & 0.00 & 0.92 \\
         &  & Non-problem & 31 & 0.85 & 0.30 & 0.00 & 0.92 \\
        \bottomrule
  	\end{tabular}
 
	\label{tab:JPAttackRes}
\end{table}

\subsection{Poisoning and Evasion Attack Interactions} \label{sec:case-study-inter}

This subsection studies the interaction between backdoor (poisoning) and adversarial malware attacks (evasion) in ML-based MD systems.
This study is inspired by previous works in the image domain~\cite{DBLP:conf/ccs/PangSZJVLLW20,DBLP:conf/nips/WengLW20}, which find these two attacks are strongly correlated and boost each other.
However, software, characterized by its discrete format and unique problem space constraints, fundamentally differs from images.
Consequently, such conclusions need to be revisited in the malware domain.
To instantiate our study, we choose AdvDroidZero~\cite{DBLP:conf/ccs/HeX0J23} as the adversarial example attack and JP~\cite{DBLP:conf/sp/YangCCPTPCW23} as the poisoning attack, because they are the SOTA methods in their respective fields.

We first train the poisoned models utilizing JP with different trigger constraints and the benign models in our dataset.
Then, we conduct the adversarial malware attack utilizing AdvDroidZero under a 10-query budget on 100 malware samples, selected uniformly randomly from malware identified by both the poisoned and the benign models.
We re-implement JP and integrate it into our evaluation framework using PyTorch~\cite{DBLP:conf/nips/PaszkeGMLBCKLGA19}, following the descriptions and configurations provided in the original paper and their open-source code.
To be specific, we use the same hyperparameters with JP to compute the trigger and set the poisoning rate as 0.001, which corresponds to 100 benign samples.
As for the target model, we use an MLP binary classifier with three hidden layers of 1,024 neurons and a dropout rate of 0.5.
As suggested by their original implementation, we use LinearSVM regularizer to select the top 10,000 features and randomly split the dataset according to the ratio of about 7:3.
For our experiment, we randomly choose 7 different malware families with various sample sizes as the targeted family.
We also extend the JP attack to various constraints, including optimizing triggers in the problem space and feature space, as well as optimizing triggers in the non-problem space.
The JP attack's effectiveness is detailed in Table~\ref{tab:JPAttackRes}, demonstrating successful backdoor injection and verifying our implementation's fidelity.

To conduct adversarial example attacks, we utilize AdvDroidZero~\cite{DBLP:conf/ccs/HeX0J23}, reconstructing its malware perturbation set by re-extracting the code perturbation in our test set.
We then select 100 malware samples, identifiable by both poisoned and benign models, and apply AdvDroidZero with the 10 query budget to generate adversarial malware examples.

Table~\ref{tab:AdvPoisonRes} reveals that the poisoned model does not consistently exhibit higher vulnerability to adversarial malware examples compared to the benign model. 
It can be observed that there are some cases in which the attack success rates in the poisoned model are not higher than the benign model whenever the constraints of the trigger.
Notably, in cases where triggers are optimized across the entire feature space, certain families like kuguo and eldorado exhibit no increase in attack success rates.
We hypothesize that such triggers might not be effective in problem space due to problem space constraints.
To validate our assumption, we truncate the trigger into the problem space and validate its backdoor effectiveness in the poisoned model.
The attack success rates in samples with the kuguo family and eldorado family of truncated triggers are both nearly 0, indicating no effect in problem space.
The attack success rates of remaining families, i.e., mobisec, tencentprotect, anydown, leadbolt, jiagu, are 0.250, 0.238, 0.438, 0.114, 0.208, respectively, indicating their effectiveness in problem space.

To further analyze the impact of problem space constraints, we take into consideration the cases of constraining the trigger in problem space.
In this case, about half of the families show no increase in attack success rates, indicating that a straightforward combination of poisoning and adversarial example attacks does not necessarily enhance evasion effectiveness.
This can be attributed to the discrete nature of the feature spaces, where the adversarial example attacks may struggle to find an optimal combination of perturbations to form the trigger exactly.

Additionally, we explore the scenario where the trigger is restricted to the non-problem space.
In this case, the injected trigger cannot be realized in the problem space, meaning that it will not introduce backdoor vulnerabilities in practice.
In this setting, 3 out of 7 families show a significantly higher attack success rate, possibly due to the interplay between non-problem and problem space features.
Conversely, 3 families exhibit a substantially lower attack success rate.
This suggests the potential for defenders to design the defense utilizing the poisoning attack by injecting the trigger in the non-problem space to enhance the adversarial robustness of the model.

In summary, the poisoned model may not necessarily be more susceptible to adversarial examples in the malware domain, probably due to the problem space constraints and the discrete feature space.
It is even possible for the defenders to design adversarial example defense methods employing poisoning attacks by leveraging the problem space constraints.

\begin{figure*}[t!]
    \centering
    \vspace{0.2em}
    \begin{subfigure}[t]{0.32\textwidth}
        \centering
        \includegraphics[width=\textwidth]{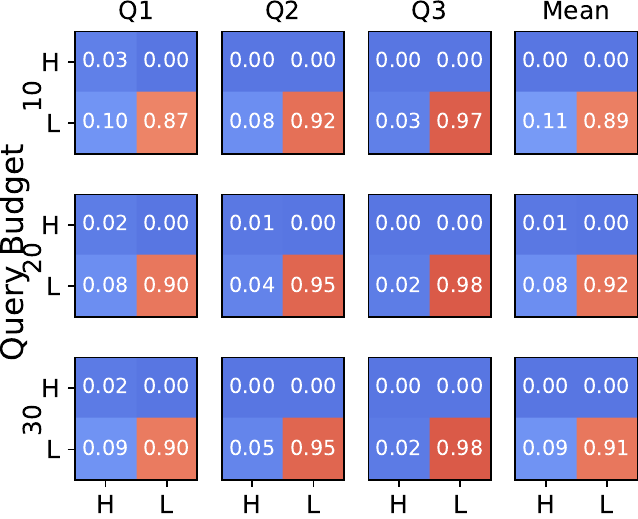}
        \caption{Approx-TCE, Drebin}
    \end{subfigure}
    \hfill
    \begin{subfigure}[t]{0.30\textwidth}
        \centering
        \includegraphics[width=\textwidth]{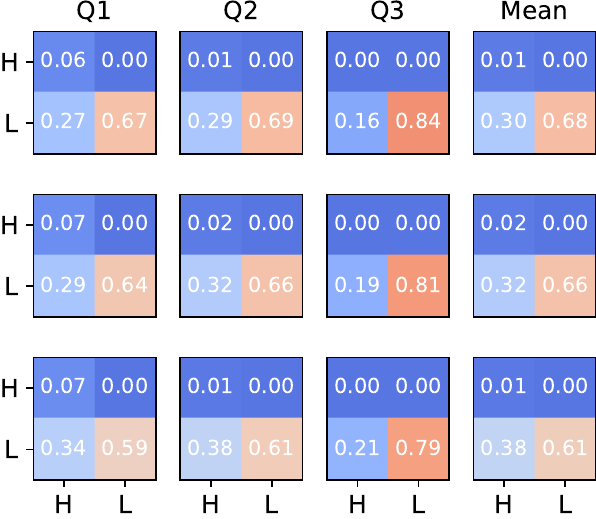}
        \caption{ICE, Drebin}
    \end{subfigure}%
    \hfill
    \begin{subfigure}[t]{0.35\textwidth}
        \centering
        \includegraphics[width=\textwidth]{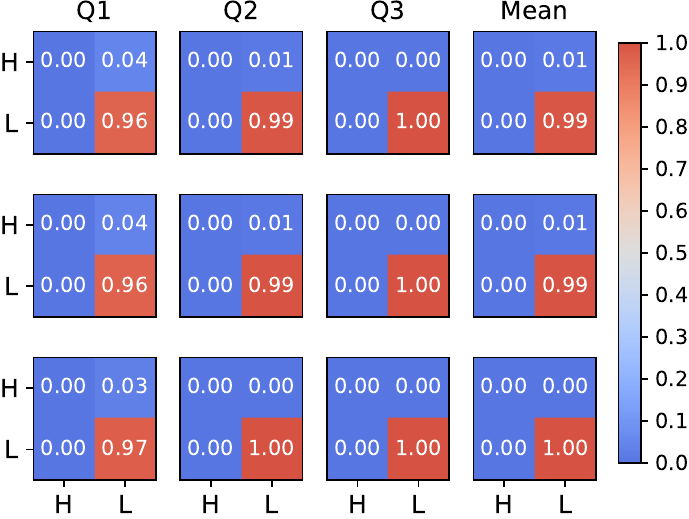}
        \caption{CCE, Drebin}

    \end{subfigure}%

    \begin{subfigure}[t]{0.32\textwidth}
        \centering
        \includegraphics[width=\textwidth]{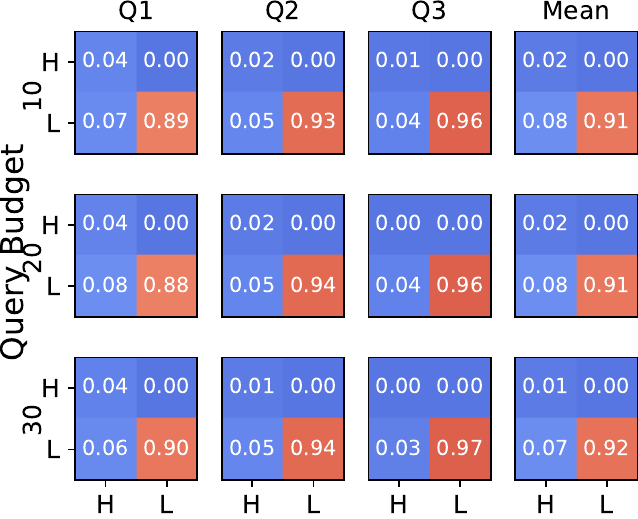}
        \caption{Approx-TCE, APIGraph}
    \end{subfigure}%
    \hfill
    \begin{subfigure}[t]{0.30\textwidth}
        \centering
        \includegraphics[width=\textwidth]{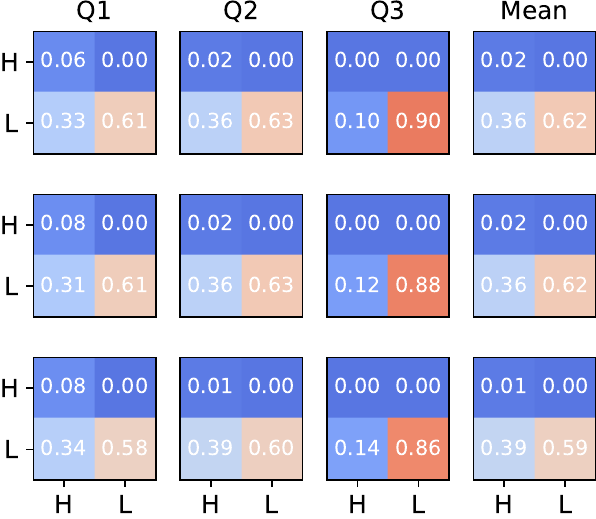}
        \caption{ICE, APIGraph}
    \end{subfigure}%
    \hfill
    \begin{subfigure}[t]{0.35\textwidth}
        \centering
        \includegraphics[width=\textwidth]{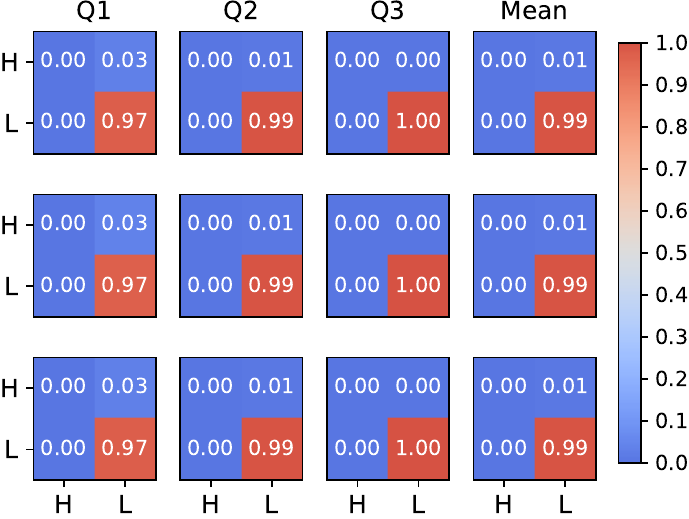}
        \caption{CCE, APIGraph}
    \end{subfigure}%

    \caption{The detection performance of TRANSCENDENT (three instantiations: Approx-TCE, ICE and CCE) against adversarial malware examples generated by AdvDroidZero with different query budget (10, 20 and 30) on different ML-based MD systems (Drebin, APIGraph). Given a predefined threshold (25\%, 50\% and 75\%, denoted by Q1, Q2 and Q3, respectively), an input is classified  based on algorithm confidence (High / Low) and credibility (High / Low). In each heatmap,  the x-axis represents the algorithm confidence, the y-axis represents the algorithm credibility, and each region represents the percentage of samples.}
    \label{fig:CDAgainstAE}
\end{figure*}

\subsection{Connections between Evasion and Concept Drift Detections}
\label{sec:case-study-intra}

Recall that concept drift detection, which aims to identify the distribution shift in the input data, is strongly similar to the detection of adversarial examples.
However, whether these methods can be unified is in question.
Therefore, this subsection studies the interaction between evasion attacks and concept drift detection in ML-based MD systems, particularly whether concept drift detection methods can detect adversarial malware examples.
We choose TRANSCENDENT~\cite{DBLP:conf/sp/BarberoPPC22} as the concept drift detection algorithm and AdvDroidZero~\cite{DBLP:conf/ccs/HeX0J23} as the adversarial example attack algorithm, due to their SOTA performance and practical applicability.

TRANSCENDENT~\cite{DBLP:conf/sp/BarberoPPC22} leverages conformal prediction theory to assess each test sample, offering two key metrics: algorithm credibility and algorithm confidence.
It then determines thresholds for these metrics, considering samples with metrics below these thresholds as having low algorithm credibility or algorithm confidence.
As TRANSCENDENT rejects concept drift samples with low algorithm credibility or low algorithm confidence, this policy requires adaptation for effective adversarial malware detection because it involves a broader range of cases, such as samples being similar to more classes.
To adapt it to detect adversarial malware examples, we only consider the samples with both low algorithm credibility and low algorithm confidence as the detected adversarial malware example, indicating the sample seems to be more similar to another label.
We re-implement the three conformal evaluators: approximate Transductive Conformal Evaluator (approx-TCE), Inductive Conformal Evaluator (ICE), and Cross-Conformal Evaluator (CCE) proposed in the TRANSCENDENT and integrate them into our evaluation framework.
Furthermore, we use different thresholds based on quartiles and consider only the correct samples to compute the threshold.
Notably, our implementation follows the latest version of TRANSCENDENT\footnote{The version can be accessed via the GitHub: \url{https://github.com/s2labres/transcendent-release}.}, which rectifies the data snooping error~\cite{DBLP:conf/uss/ArpQPWPWCR22}.

\begin{table}[t]
    \setlength{\abovecaptionskip}{0pt}
	\caption{Detection performance of the selected MD-based MD systems and the number of adversarial malware examples successfully generated by AdvDroidZero under different query budgets.}
 
	\begin{tabular}[centering,width=0.5*\linewidth]{@{}C{1.5cm}C{1cm}C{1cm}C{0.9cm}C{0.6cm}C{1.2cm}@{}}
		\toprule
        \multirow{2}{*}{Methods} & \multicolumn{2}{c}{Performance} & \multicolumn{3}{c}{Query Budget} \\ \cmidrule(l){2-3} \cmidrule(l){4-6}
         & TPR & F1 & 10 & 20 & 30 \\
		\midrule
        Drebin & 0.91 & 0.93 & 502 & 586 & 600 \\
        APIGraph & 0.91 & 0.92 & 392 & 503 & 551 \\
        \bottomrule
  	\end{tabular}
 
	\label{tab:CDModelRes}
\end{table}

For generating adversarial malware examples, we select Drebin~\cite{DBLP:conf/ndss/ArpSHGR14} and APIGraph~\cite{DBLP:conf/ccs/ZhangZZDCZZY20} as target ML-based MD systems due to their various kinds of feature types, including syntax features and semantic features.
To ensure fidelity to their original implementations, we strictly follow the configurations provided in their respective publications.
Therefore, we employ SVM with the linear kernel as the target classifier, which is consistent with the previous work~\cite{DBLP:conf/ccs/HeX0J23}.
To simulate a real-world ML-based MD system, we utilize the time-aware split for training.
The detection performance of these methods is shown in Table~\ref{tab:CDModelRes}.
For instance, the TPR values of Drebin and APIGraph are 0.914 and 0.912, respectively, indicating high detection effectiveness.
Then, we randomly select 1,000 malware samples that can be successfully identified by the selected methods and employ AdvDroidZero with different query budgets to generate adversarial malware examples.
The actual number of successfully generated adversarial malware examples is presented in Table~\ref{tab:CDModelRes}.
Subsequently, the three conformal evaluators are employed to detect these adversarial examples.

As depicted in Figure~\ref{fig:CDAgainstAE}, TRANSCENDENT exhibits high true positive rates across various settings, confirming its effectiveness in identifying adversarial malware examples.
This success is likely due to the adversarial generation process, which typically targets vulnerable features, reducing the confidence in the malware label.
These features differ from the features that make the malware samples similar to samples in the training set.
Therefore, TRANSCENDENT can be effective in detecting adversarial malware examples.
Furthermore, the CCE achieves the most strong detection effectiveness among the three conformal evaluators, which achieves nearly 100\% of the true positive rate.
This may be attributed to the ensemble design in the CCE increasing the difficulty for the adversarial malware examples.
In practice, the false positive rate, influenced by true concept drift samples, is also a critical consideration.
As the false positive rate is introduced because of the true concept drift samples, the concept drift detection methods can be deployed in the initial days when there are fewer concept drift samples.
It also opens up the future design to adapt concept drift detection methods in adversarial malware detection by distinguishing the adversarial malware examples and the true concept drift samples.

In summary, concept drift detection methods in the malware domain show high true positive rates in detecting adversarial malware examples.
To minimize false positives caused by real concept drift samples, it is advisable to deploy these detection methods early on, reducing the likelihood of encountering such samples.

\section{Future Directions}

Building upon our stage-based taxonomy, we have identified significant technical advancements and limitations in current attack and defense proposals about the security risks of ML-based MD systems.
However, our exploration is merely the beginning of a more extensive journey.
This section outlines potential future research directions, aiming to deepen our understanding of security risks and develop secure and reliable ML-based MD systems.
We propose these directions within our stage-based framework, dividing them into two primary categories: inter-stage future directions and intra-stage future directions.
We remark that these directions are by no means exhaustive and only serve as a starting point for future research.

\subsection{Inter-Stage Future Directions}

\paragraphbe{Understanding the Attack Connections}
In all current attack proposals within ML-based MD systems, a prevailing assumption is that a singular attack vector is in play~\cite{DBLP:conf/sp/YangCCPTPCW23,DBLP:conf/ccs/HeX0J23,DBLP:conf/sp/PierazziPCC20}.
However, this simplification might not accurately capture the complexity of real-world threats, where attackers are frequently motivated to amplify their impact by simultaneously deploying multiple types of attacks.
For instance, the IMC attack~\cite{DBLP:conf/ccs/PangSZJVLLW20} leverages both input perturbation and model poisoning techniques to enhance its attack effectiveness in the image domain.
This example is an inspiration to investigate the potential for similar multifaceted attacks within the malware domain, as showcased in Section~\ref{sec:case-study-inter}.
Consequently, there is a clear opportunity for further research to deepen our understanding of stronger attacks that combine multiple attack vectors.

\paragraphbe{Holistic Defense Mechanism}
As shown in Table~\ref{tab:representativeworks}, each stage of ML-based MD systems exhibits susceptibility to specific attack types, with some attacks not having countering defenses at the same stage.
Furthermore, in practice, defenders aim to secure ML-based MD systems against all potential security threats, requiring that the defense mechanism to incorporate defense methods across different stages against various attacks.
Additionally, minimizing the deployment costs of such defenses is a crucial consideration.
Therefore, one promising direction is the development of a holistic defense mechanism.
Such approach would entail a dynamic incorporation of existing defense methods tailored to address the full range of potential attacks across different stages of ML-based MD systems.

\subsection{Intra-Stage Future Directions}

\paragraphbe{Robust Dataset Preprocessing Method}
In ML-based MD systems, the availability of massive software datasets from open-source platforms presents both opportunities and challenges.
The constraints of computational cost and labeling demand necessitate efficient dataset preprocessing methods.
Additionally, these methods must accurately reflect the real-world temporal and spatial distribution of malware data to avoid experimental bias.
Thus, preprocessing methods that are robust against both time and space constraints are required yet missing.
Moreover, open-source datasets are potentially vulnerable to poisoning attacks.
Currently, such attacks often overlook the impact of dataset preprocessing methods.
Evaluating the effectiveness of poisoning attacks in the context of robust preprocessing will provide deeper insights into the practical availability risks associated with ML-based MD systems.
This understanding is instrumental in developing and refining preprocessing methods to mitigate such risks effectively.
This direction would enhance the robustness of ML-based MD systems against data poisoning and ensure their effectiveness and reliability in real-world dynamic scenarios.

\paragraphbe{Practical Problem Space Solutions}
To understand the semantics of software data, ML-based MD systems rely on program analysis tools to map the input data from the problem space to the feature space.
From the attack side, the attackers need to perform problem space modification to conduct practical attacks.
While there are some existing problem space modification solutions, they have many restrictions, such as the dependency on try-catch mechanisms, thereby constraining their ability to fully exploit problem space vulnerabilities.
This limitation highlights the need for more sophisticated and flexible problem space modification techniques that can uncover and leverage a broader range of vulnerabilities.
From the defensive perspective, there are inherent challenges in fully understanding the program semantics of malware data, often due to limitations in program analysis tools.
For instance, static analysis can suffer from the path explosion problem, hindering comprehensive analysis.
Additionally, the diverse nature of malware necessitates various program analysis methods for effective detection across different scenarios.
As such, designing a better mapping from the problem space to the feature space would improve the detection effectiveness and robustness of ML-based MD systems.

\paragraphbe{Reliable Model Building Methods}
As discussed in Section~\ref{sec:sr3}, attacks at Stage 3 of the ML-based MD systems remain largely unexplored.
Drawing parallels from the image domain~\cite{bagdasaryan2021blind}, one potential security risk in the malware domain involves attackers injecting vulnerabilities into ML models through open-source ML libraries.
This emerging threat underscores the need to explore such security risks within the malware domain.
Beyond awareness of code reliability, most of the existing robust model architectures and model training methods lack provable robustness guarantees.
Although certified training methods have started to gain interest in the malware domain~\cite{DBLP:conf/uss/Chen0SJ20,DBLP:conf/ccs/GibertZL23,huang2023rs}, they focus on limited input specifications, e.g., norm-based input constraints, while the malware domain requires more sophisticated specifications due to diverse discrete feature spaces and different semantic constraints.
Future research may focus on developing novel metrics for constraints more tailored to the malware domain, alongside enhancing the adaptability and applicability of certified training methods.

\paragraphbe{Unified Anomaly Input Detection}
At Stage 4 of ML-based MD systems, the defender can potentially unify concept drift detection methods and adversarial malware samples detection, as showcased in Section~\ref{sec:case-study-intra}.
Therefore, this raises a critical challenge for adversarial example attacks, which must now evade both malware detection and concept drift detection.
This complexity also necessitates a more thorough exploration of the efficacy of concept drift detection methods in identifying adversarial malware, which can be pivotal in enabling defenders to develop more effective methods to detect adversarial malware.
Moreover, the scope of anomaly detection may be further extended to include other types of abnormal inputs, such as backdoor input samples.
Such possibility calls for development of a comprehensive, unified anomaly input sample detection method.
By integrating different forms of anomaly detection into a unified framework, defenders can significantly improve the utility and robustness of their security measures.

\section{Related Surveys}
\label{sec:relatedwork}

Adversarial ML has recently gathered significant attention in various domains, such as image, text, audio, etc.
In the malware detection domain, prior surveys~\cite{DBLP:journals/csur/MaiorcaBG19,DBLP:journals/comsur/YanRWSZY23,DBLP:journals/csur/LiLYX23,DBLP:journals/compsec/LingWZQDCQWJLWW23,DBLP:journals/tissec/DemetrioCBLAR21} summarize the existing attack and defense proposals in ML-based MD systems based their assumptions and methodologies.
Maiorca \textit{et al}.~\cite{DBLP:journals/csur/MaiorcaBG19} categorized the threats specifically targeted against ML-based PDF malware detectors based on the threat model.
Yan \textit{et al}.~\cite{DBLP:journals/comsur/YanRWSZY23} summarized the existing adversarial example attack and defense methods for ML-based MD methods based on the approaches.
Li \textit{et al}.~\cite{DBLP:journals/csur/LiLYX23} proposed a concept systematization framework based on the attributes of threat models.
Ling \textit{et al}.~\cite{DBLP:journals/compsec/LingWZQDCQWJLWW23} surveyed the adversarial example attack methods on ML-based Windows PE malware detection methods based on the adversary knowledge.

However, these surveys do not provide a holistic and comprehensive security analysis framework from the practical system development viewpoint to understand the landscape and new potential attacks and defense of ML-based MD systems.
For instance, they may concentrate on one kind of threat, e.g., adversarial example attacks~\cite{DBLP:journals/comsur/YanRWSZY23,DBLP:journals/compsec/LingWZQDCQWJLWW23}.
Although such concentration may help to design better adversarial example attack methods, it hardly helps to understand connections between security risks, e.g., the connections between attacks.
Additionally, they may not stand in the viewpoint of the ML-based MD system pipeline, which limits their practical implications.

\section{Conclusion}

This paper provides a stage-based taxonomy for a holistic understanding of the security risks of ML-based MD systems. 
The stage-based taxonomy is derived by dissecting an ML-based MD system into its operational stages.
Leveraging the stage-based taxonomy, we summarize the technical progress of the related attacks and defenses with their limitations in each stage.
In addition, we provide two case studies from the perspective of inter-stage and intra-stage with empirical insights. 
Furthermore, based on the taxonomy and analyses, we discuss the potential inter-stage and intra-stage future directions.

\bibliographystyle{ACM-Reference-Format}
\bibliography{reference.bib}

\end{document}